\documentclass[secnumarabic,aps,prd,floatfix,nofootinbib,twocolumn,showkeys,10pt,superscriptaddress]{revtex4-1}
\usepackage{amsmath,amssymb,amsfonts}
\usepackage{graphicx}
\usepackage{xcolor}
\usepackage[colorlinks=true,linkcolor=blue,citecolor=blue,urlcolor=blue]{hyperref}

\def\Lm{\mathcal{L}_m}

\newcommand{\ns}{n_{\rm s}}

\begin{document}

\title{Inflation in $f(Q,\mathcal{L}_m)$ gravity: equivalence to $k$-inflation and ghost instabilities}

\author{Maryam Shiravand}
\email{ma\_shiravand@kntu.ac.ir}
\affiliation{Department of Physics, K.N. Toosi University of Technology, P.O. Box 15875-4416, Tehran, Iran}

\author{Saeed Fakhry}
\email{saeed.fakhry@uv.es}
\affiliation{Departamento de Astronom\'{\i}a y Astrof\'{\i}sica, Universitat de Val\`encia, Avinguda Vicent Andr\'es Estell\'es 19, 46100 Burjassot (Val\`encia), Spain}

\author{Javad T. Firouzjaee}
\email{firouzjaee@kntu.ac.ir}
\affiliation{Department of Physics, K.N. Toosi University of Technology, P.O. Box 15875-4416, Tehran, Iran}

\author{Ali Tizfahm}
\email{alitizfahm@aut.ac.ir}
\affiliation{Department of Physics, K.N. Toosi University of Technology, P.O. Box 15875-4416, Tehran, Iran}

\begin{abstract}
We study slow-roll inflation in $f(Q,\mathcal{L}_m)$ gravity, focusing on whether non-minimal matter--geometry couplings can support viable inflation. We show that the separable class linear in $Q$, $f(Q,\mathcal{L}_m)=-\alpha Q+g(\mathcal{L}_m)$, is equivalent, up to a boundary term, to General Relativity minimally coupled to a $k$-essence scalar field. Thus, it introduces no additional gravitational degrees of freedom, and its inflationary predictions coincide with those of the corresponding $k$-inflation model. For the separable models considered, the observables depend on a single effective parameter combination, yielding a nonzero lower bound on the tensor-to-scalar ratio. While these models may marginally satisfy Planck 2018 constraints alone, they are strongly disfavored by the combined Planck 2018, baryon acoustic oscillation (BAO), and BICEP/Keck 2018 (BK18) constraints. This marginal agreement also requires a sizeable negative effective vacuum-energy contribution, driving the expanding background to $H=0$ shortly after inflation. For representative non-separable models with genuine $Q$--$\mathcal{L}_m$ couplings, including terms proportional to $\gamma QV$ and $\gamma QX$, we derive the quadratic perturbation action and find that two additional scalar degrees of freedom propagate on regular rolling backgrounds satisfying the relevant non-degeneracy conditions. Under these conditions, the kinetic matrix has at least one negative eigenvalue, indicating a ghost instability. Hence, background-level observational predictions alone cannot establish the physical viability of these models. Overall, for the classes considered, inflation in $f(Q,\mathcal{L}_m)$ gravity either reduces to $k$-inflation in General Relativity or exhibits ghost instabilities on regular, non-degenerate backgrounds.
\end{abstract}


\maketitle

\section{Introduction}\label{sec:intro}


Cosmological inflation, an early epoch of accelerated cosmic expansion, successfully resolves the horizon, flatness, and unwanted relic problems of the standard hot Big Bang scenario while providing a causal, quantum-mechanical origin for the primordial perturbations that seed large-scale structure \citep{starobinsky1980new,guth1981inflationary,linde1982new,albrecht1982cosmology}. Its canonical realizations, based on a single minimally coupled scalar field slowly rolling down an effective potential within the framework of General Relativity (GR), yield sharp predictions for the scalar spectral index $\ns$, the tensor-to-scalar ratio $r$, and primordial non-Gaussianities, enabling high-precision cosmic microwave background (CMB) measurements to discriminate among competing models \citep{akrami2020planck,2021PhRvL.127o1301A,martin2014encyclopaedia}. In particular, canonical large-field monomial potentials $V(\phi)\propto\phi^n$ that underpin chaotic inflation \citep{linde1983chaotic} are now severely disfavored or excluded, primarily because their predicted amplitude of primordial gravitational waves exceeds the tight empirical bound $r<0.036$ established by the joint Planck 2018, BICEP/Keck 2018 (BK18), and baryon acoustic oscillation (BAO) data \citep{2021PhRvL.127o1301A}.

To reconcile inflationary predictions with such tight observational bounds, extensive efforts have explored extended theories of gravity \citep{de2010fr,sotiriou2010fr,nojiri2011unified,capozziello2011extended,clifton2012modified}, ranging from Riemannian curvature modifications, mimetic gravity \citep{2015arXiv150707726H}, and non-minimal matter--curvature couplings such as $f(R,\mathcal{L}_m)$ \citep{harko2010frlm} and $f(R,T)$ (supplemented optionally by topological or parity-violating corrections \citep{2026arXiv260720654S}), to alternative geometric descriptions. A prominent alternative is symmetric teleparallel gravity, wherein gravitational interactions are described by the non-metricity tensor $Q_{\alpha\mu\nu}$ of a flat, torsion-free connection rather than spacetime curvature \citep{nester1999symmetric,jimenez2018coincident,jimenez2018teleparallel,beltran2019geometrical}. In parallel to pure geometric $f(Q)$ theories \citep{jimenez2018coincident,jimenez2020cosmology,anagnostopoulos2021fQ} and $f(Q,T)$ formulations \citep{xu2019fq,Shiravand:2022ccb,Shiravand:2024ayw}, matter--geometry couplings have recently been extended to $f(Q,\mathcal{L}_m)$ gravity, where non-metricity couples directly to the matter Lagrangian density $\mathcal{L}_m$ \citep{2025PDU....5002092H}. This framework has motivated numerous cosmological investigations of both early-universe inflation and late-time acceleration \citep{2024PDU....4601614M,2024JHEAp..44..164M,2024arXiv241005335S,2025A&C....5200956G,2025EPJC...85..376M,2025PhLB..86639506M,2025PDU....4801829M,2025Ap&SS.370...92S}.

However, virtually all existing inflationary studies in $f(Q,\mathcal{L}_m)$ gravity rely exclusively on background modified Friedmann equations. Such background-only analyses suffer from two critical shortcomings: first, whenever the action is separable and linear in $Q$, the non-minimal matter--geometry coupling can be recast as a modification of the matter sector, introducing no genuinely new gravitational degrees of freedom beyond standard GR; second, nonlinear extensions of symmetric teleparallel gravity can exhibit nontrivial constraint structures and potentially pathological scalar perturbation sectors, even when analyzed in the coincident gauge ($\Gamma^\alpha{}_{\mu\nu}=0$) \citep{jimenez2020cosmology,gomes2024pathological,jimenez2026nonviability,dambrosio2022degrees}. Indeed, spatially flat cosmological backgrounds in pure $f(Q)$ gravity have been associated with strong-coupling problems or ghost instabilities \citep{gomes2024pathological}, while general theories of the form $f(Q,\chi,(\partial\chi)^2)$ generically propagate ghost modes under the conditions discussed in the literature \citep{jimenez2026nonviability}. Since the inflaton matter Lagrangian takes the form $\mathcal{L}_m=X-V(\phi)$, with $X\equiv-\frac{1}{2}g^{\mu\nu}\partial_\mu\phi\partial_\nu\phi$, the $f(Q,\mathcal{L}_m)$ framework constitutes a particular subclass of theories in which non-metricity is coupled to both the inflaton kinetic and potential sectors. This makes a dedicated perturbative analysis essential, rather than allowing the stability properties to be inferred solely from the background equations.

The central novelty and core thesis of this work is that \textit{background-level inflationary viability does not guarantee perturbative consistency in non-separable $f(Q,\mathcal{L}_m)$ gravity}. To demonstrate this, we systematically examine the theoretical structure of $f(Q,\mathcal{L}_m)$, delineate the limitations of background-only slow-roll analyses, and perform a degrees-of-freedom and perturbative stability analysis for the representative model classes considered here. Specifically, we demonstrate that the separable class linear in $Q$, described by $f(Q,\mathcal{L}_m)=-\alpha Q+g(\mathcal{L}_m)$, is exactly equivalent, up to a boundary term, at the action level, to General Relativity minimally coupled to a $k$-essence scalar field \citep{armendariz1999kinflation,garriga1999perturbations,chiba2000kinflation}. This equivalence implies that no additional gravitational degrees of freedom are introduced by this class. For the separable models considered in this work, we find that the tensor-to-scalar ratio is bounded from below, with $r>0.066$ for $N\leq60$. Consequently, although these models may exhibit marginal compatibility with Planck 2018 constraints considered alone, they are strongly disfavored by the combined Planck 2018, BK18, and BAO data. Moreover, this marginal agreement requires a sizeable negative effective vacuum-energy contribution, which drives the expanding background toward $H=0$ after inflation. Finally, for representative non-separable models featuring genuine $Q\text{--}\mathcal{L}_m$ interactions, we compute the quadratic scalar perturbation action and show that, on regular rolling backgrounds satisfying the relevant non-degeneracy conditions, two additional scalar degrees of freedom propagate. Furthermore, under the corresponding conditions, the kinetic matrix possesses at least one negative eigenvalue, signaling a ghost instability. Therefore, background-level observational predictions alone cannot establish the physical viability of these coupled models.

The remainder of this paper is structured as follows. In Sec.~\ref{sec:sr}, we briefly recapitulate the baseline slow-roll dynamics in GR. Section~\ref{sec:fqlm} outlines the mathematical foundation of $f(Q,\mathcal{L}_m)$ gravity and its background cosmology. In Sec.~\ref{sec:separable}, we analyze the separable model class, demonstrating its exact correspondence to $k$-inflation and its observational constraints. Section~\ref{sec:coupled} is devoted to the non-separable coupled class, where we compute the quadratic perturbation action and establish the ghost instability results for representative interaction models on regular rolling backgrounds. We discuss broader implications and theoretical caveats in Sec.~\ref{sec:discussion}, and provide our concluding remarks in Sec.~\ref{sec:conclusions}. Relevant analytical derivations and perturbation matrices are detailed in the Appendices. Throughout this work, we adopt natural units where $\hbar=c=1$. In the context of General Relativity (Sec.~\ref{sec:sr}), we set the reduced Planck mass $M_{\rm Pl}=1$. In the $f(Q,\mathcal{L}_m)$ framework, the gravitational normalization is determined by $\alpha$, with $M_{\rm Pl}^2=\alpha$ on the standard-sign branch $\alpha>0$. Where appropriate, we use this normalization to simplify the numerical analysis.

\section{Slow-roll inflation in General Relativity}
\label{sec:sr}

For a canonical scalar field $\phi$ endowed with a potential $V(\phi)$ in a spatially flat Friedmann--Lema\^{i}tre--Robertson--Walker
(FLRW) spacetime, described by the line element
\begin{equation}
\mathrm{d}s^2=-\mathrm{d}t^2
+a^2(t)\delta_{ij}\mathrm{d}x^i\mathrm{d}x^j,
\end{equation}
the governing Friedmann and Klein--Gordon equations read
\begin{align}
3H^2 &= \frac{\dot\phi^2}{2}+V,
\qquad
\dot H=-\frac{\dot\phi^2}{2},
\label{eq:GRfried}\\
\ddot\phi&+3H\dot\phi+V'=0.
\label{eq:GRkg}
\end{align}
Here and throughout this section, we use the reduced Planck
mass convention $M_{\rm Pl}=1$.

The Hubble-flow parameters are defined as
$\epsilon_1\equiv-\dot H/H^2$ and
$\epsilon_{n+1}\equiv
\mathrm{d}\ln|\epsilon_n|/\mathrm{d}\mathcal{N}$, where $\mathcal{N}\equiv\ln a$ denotes the number of
e-folds increasing with cosmic time
\citep{martin2014encyclopaedia}. Denoting
$\epsilon\equiv\epsilon_1$, the second slow-roll parameter
$\eta$ is defined by
\begin{equation}
\label{eq:etadef}
\eta\equiv 2\epsilon-\frac{\epsilon_2}{2}
=-\frac{\dot H}{H^2}
-\frac{\ddot H}{2H\dot H}
=\epsilon-\frac{\ddot\phi}{H\dot\phi},
\end{equation}
where the last equality is exact for a canonical scalar field. At leading order in slow roll, the potential slow-roll parameters are given by
\begin{equation}
\epsilon\simeq\frac{1}{2}
\left(\frac{V'}{V}\right)^2,
\qquad
\eta\simeq\frac{V''}{V}.
\end{equation}
The corresponding leading-order expressions for the scalar spectral index and tensor-to-scalar ratio are
\begin{equation}
\label{eq:obsGR}
n_s=1-6\epsilon+2\eta,
\qquad
r=16\epsilon.
\end{equation}

The number of e-folds remaining between the horizon exit of a pivot scale and the end of inflation is defined as
\begin{equation}
\label{eq:NGR}
N \equiv\int_t^{t_{\rm end}}H\,\mathrm{d}t
=-\int_{\phi_{\rm end}}^{\phi_*}
\frac{H}{\dot\phi}\,\mathrm{d}\phi
\simeq
\int_{\phi_{\rm end}}^{\phi_*}
\frac{V}{V'}\,\mathrm{d}\phi,
\end{equation}
where $\phi_*$ denotes the field value at horizon exit and the end of inflation is determined by
$\epsilon(\phi_{\rm end})=1$.

For a quadratic potential
$V(\phi)=\frac{1}{2}m^2\phi^2$, these expressions yield,
at leading order in slow roll,
\begin{equation}
n_s\simeq1-\frac{2}{N},
\qquad
r\simeq\frac{8}{N}.
\end{equation}
Thus, for $N=50$--$60$, one obtains $r\simeq0.13$--$0.16$, which is incompatible with the
observational upper bound on the tensor-to-scalar ratio
\citep{2021PhRvL.127o1301A}.

For a non-canonical scalar field described by a Lagrangian
$P(X,\phi)$, where
$X\equiv-\frac{1}{2}g^{\mu\nu}
\partial_\mu\phi\partial_\nu\phi$, scalar perturbations propagate with the sound speed
\begin{equation}
c_s^2=\frac{P_X}{P_X+2XP_{XX}},
\end{equation}
where subscripts denote derivatives with respect to $X$. At leading order in the slow-roll and sound-speed
variation parameters, the observable spectral quantities are modified to \citep{garriga1999perturbations}
\begin{equation}
\label{eq:obsK}
n_s-1=-2\epsilon_1-\epsilon_2-s,
\qquad
r=16c_s\epsilon_1,
\qquad
s\equiv\frac{\dot c_s}{Hc_s}.
\end{equation}
The canonical GR expressions are recovered in the limit
$c_s=1$, for which $s=0$.

\section{$f(Q,\mathcal{L}_m)$ gravity}
\label{sec:fqlm}


In symmetric teleparallel geometry, the connection is flat and torsion-free, and gravity is encoded in the non-metricity tensor
$Q_{\alpha\mu\nu}\equiv\nabla_\alpha g_{\mu\nu}$
\citep{jimenez2018coincident}. Defining its two traces as
$Q_\alpha\equiv Q_\alpha{}^\mu{}_\mu$ and
$\tilde Q_\alpha\equiv Q^\mu{}_{\alpha\mu}$, the non-metricity scalar is given by
\begin{equation}
\label{eq:Qdef}
Q=
\frac{1}{4}Q_{\alpha\mu\nu}Q^{\alpha\mu\nu}
-\frac{1}{2}Q_{\alpha\mu\nu}Q^{\mu\alpha\nu}
-\frac{1}{4}Q_\alpha Q^\alpha
+\frac{1}{2}Q_\alpha\tilde Q^\alpha.
\end{equation}
We adopt the convention in which $Q=6H^2$ on a spatially flat FLRW background in the coincident gauge. With this
convention, the non-metricity scalar and the Ricci scalar $\mathring R$ of the Levi-Civita connection are related by
\begin{equation}
\label{eq:QR}
\sqrt{-g}\,Q
=-\sqrt{-g}\,\mathring R
+\partial_\mu\left(\sqrt{-g}\,\mathcal{B}^\mu\right),
\end{equation}
where $\mathcal{B}^\mu$ is a boundary vector constructed from the non-metricity traces. Since the connection is
flat and torsion-free, it can locally be set to zero by an appropriate coordinate choice. This coordinate choice defines the coincident gauge, in which
$\Gamma^\alpha{}_{\mu\nu}=0$ and
$Q_{\alpha\mu\nu}=\partial_\alpha g_{\mu\nu}$.

The action of $f(Q,\mathcal{L}_m)$ gravity is given by \citep{2025PDU....5002092H}
\begin{equation}
\label{eq:action}
S=\frac{1}{2}\int\mathrm{d}^4x\,\sqrt{-g}\,
f(Q,\mathcal{L}_m).
\end{equation}

Throughout this work, we consider the inflaton as the matter source and adopt the canonical scalar-field Lagrangian
\begin{equation}
\mathcal{L}_m=X-V(\phi),
\qquad
X\equiv-\frac{1}{2}g^{\mu\nu}
\partial_\mu\phi\partial_\nu\phi.
\end{equation}
For a homogeneous scalar field in a spatially flat FLRW background, this choice gives
$\mathcal{L}_m=p=X-V(\phi)$ and
$\rho=X+V(\phi)$, where $p$ and $\rho$ denote the
pressure and energy density, respectively.

For a spatially flat FLRW spacetime in the coincident gauge with Cartesian spatial coordinates, the background cosmological equations take the form
\citep{2025PDU....5002092H}
\begin{align}
3H^2
&=\frac{1}{4f_Q}
\left[
f-f_{\mathcal{L}_m}
(\rho+\mathcal{L}_m)
\right],
\label{eq:F1}\\
3H^2+2\dot H
&=\frac{1}{4f_Q}
\left[
f+f_{\mathcal{L}_m}
(\rho+2p-\mathcal{L}_m)
\right]
-2H\frac{\dot f_Q}{f_Q},
\label{eq:F2}
\end{align}
where $f_Q\equiv\partial f/\partial Q$ and
$f_{\mathcal{L}_m}\equiv
\partial f/\partial\mathcal{L}_m$.
These expressions are understood on regular branches where $f_Q\neq0$. We have also checked that, for the
models considered in this work, the background equations are reproduced by varying the corresponding reduced
minisuperspace action with respect to the lapse and scale factor, with the lapse retained during variation and set to unity afterwards.

The last term in Eq.~\eqref{eq:F2} represents an
additional contribution associated with the time
dependence of $f_Q$. During an expanding phase with
$H\neq0$ and $f_Q\neq0$, this term vanishes when
$f_Q$ is constant along the background evolution.

\section{Separable models: equivalence to $k$-inflation}\label{sec:separable}

\subsection{Equivalence theorem}
Consider the class of models given by
\begin{equation}\label{eq:sepclass}
f(Q,\mathcal{L}_m)=-\alpha Q+g(\mathcal{L}_m),
\qquad \alpha=\text{const}\neq 0,
\end{equation}
where $g$ is an arbitrary function. Using Eq.~\eqref{eq:QR} and discarding the total-derivative boundary term, the action
\eqref{eq:action} becomes
\begin{equation}\label{eq:sepGR}
S=\int \mathrm{d}^4x\sqrt{-g}
\left[
\frac{\alpha}{2}\mathring R
+\frac{1}{2}g\big(X-V(\phi)\big)
\right].
\end{equation}

Since $f_Q=-\alpha$ is constant and $\mathcal{L}_m$ does not depend on the symmetric teleparallel connection,
the connection field equation is identically satisfied. Consequently, the connection does not contribute to the
dynamics of this class. Equation~\eqref{eq:sepGR} is therefore, up to a boundary term, exactly equivalent at
the action level to General Relativity minimally coupled to a $k$-essence scalar field.

The physical interpretation depends on the sign of $\alpha$. For $\alpha>0$, the action has the standard gravitational
sign, with
\begin{equation}
M_{\rm Pl}^2=\alpha,
\qquad
P(X,\phi)=\frac{1}{2}g(X-V).
\end{equation}
For $\alpha<0$, multiplying the entire action by $-1$ gives the standard gravitational sign, with
\begin{equation}
M_{\rm Pl}^2=|\alpha|,
\qquad
P(X,\phi)=-\frac{1}{2}g(X-V).
\end{equation}
Thus, defining $\sigma=\operatorname{sgn}(\alpha)$, both branches can be described by
\begin{equation}
P(X,\phi)=\frac{\sigma}{2}g(X-V),
\qquad
M_{\rm Pl}^2=|\alpha|.
\end{equation}

No additional gravitational degree of freedom is introduced by this separable class. On a regular background with a
non-degenerate scalar kinetic sector, the propagating spectrum consists of the two tensor modes of GR and one
scalar degree of freedom. The quadratic action for the curvature perturbation has a kinetic coefficient
proportional to $X(P_X+2XP_{XX})$ and a gradient coefficient proportional to $XP_X$
\citep{garriga1999perturbations}. For $X>0$, the absence of ghost and gradient instabilities requires
\begin{equation}\label{eq:stability}
\begin{aligned}
P_X+2XP_{XX}
&=\frac{\sigma}{2}
\left(g'+2Xg''\right)>0,
\\
c_s^2
&=\frac{P_X}{P_X+2XP_{XX}}
=\frac{g'}{g'+2Xg''}>0,
\end{aligned}
\end{equation}
where primes on $g$ denote derivatives with respect to its argument, evaluated at $\bar{\mathcal{L}}_m=X-V(\phi)$.
All inflationary observables therefore follow from the standard $k$-inflation formulae in Eq.~\eqref{eq:obsK}.
Consequently, this separable class does not provide a genuinely new gravitational mechanism for inflation:
its predictions can be reproduced by an appropriately chosen scalar-field Lagrangian in GR.

\subsection{Linear model}

The linear model \citep{2025PDU....5002092H} is defined by
\begin{equation}\label{eq:linmodel}
f=-\alpha Q+2\mathcal{L}_m+\beta.
\end{equation}
It belongs to the separable class discussed above, with $g(\mathcal{L}_m)=2\mathcal{L}_m+\beta$.
For the standard gravitational branch $\alpha>0$, the corresponding $k$-essence Lagrangian is
\begin{equation}
P=X-V(\phi)+\frac{\beta}{2}.
\end{equation}
Thus, the model is equivalent to GR with a canonical scalar field, Planck mass $M_{\rm Pl}^2=\alpha$, and
effective potential
\begin{equation}
U(\phi)=V(\phi)-\frac{\beta}{2}.
\end{equation}
The constant parameter $\beta$ contributes an effective cosmological constant
\begin{equation}
\Lambda_{\rm eff}=-\frac{\beta}{2\alpha}.
\end{equation}
The branch $\alpha<0$, after reversing the overall sign of the action to obtain the standard gravitational sign,
has a wrong-sign scalar kinetic term and is therefore ghostly.

For $\alpha>0$, Eqs.~\eqref{eq:F1} and \eqref{eq:F2} reduce to
\begin{equation}\label{eq:linfried}
3H^2=\frac{\dot\phi^2+2V-\beta}{2\alpha},
\qquad
\dot H=-\frac{\dot\phi^2}{2\alpha},
\end{equation}
together with the standard Klein--Gordon equation \eqref{eq:GRkg}. At leading order in slow roll, the
potential slow-roll parameters are
\begin{align}
\epsilon &\simeq
\frac{\alpha}{2}
\left(\frac{V'}{V-\beta/2}\right)^2,
\qquad
\eta \simeq
\frac{\alpha V''}{V-\beta/2},
\label{eq:linsr}\\
N &\simeq
\frac{1}{\alpha}
\int_{\phi_{\rm end}}^{\phi}
\frac{V-\beta/2}{V'}\,\mathrm{d}\phi.
\label{eq:linN}
\end{align}
Here, $\phi$ denotes the field value at horizon crossing, and $N$ is the number of $e$-folds remaining until the
end of inflation. The condition $H^2>0$ requires $V>\beta/2$ for $\alpha>0$.

\subsection{Nonlinear model}

We next consider the nonlinear model
\citep{2025EPJC...85..376M,2024PDU....4601614M}
\begin{equation}\label{eq:nlmodel}
f=-\alpha Q+(2\mathcal{L}_m)^2+\beta.
\end{equation}
For this model, Eqs.~\eqref{eq:F1} and \eqref{eq:F2} give
\begin{align}
3H^2
&=
\frac{-3\dot\phi^4+4V\dot\phi^2+4V^2+\beta}
{2|\alpha|},
\label{eq:nlfried}\\
\dot H
&=
\frac{\dot\phi^2(\dot\phi^2-2V)}
{|\alpha|},
\label{eq:nlHdot}
\end{align}
along with the modified Klein--Gordon equation
\begin{equation}\label{eq:nlkg}
\left(2V-3\dot\phi^2\right)\ddot\phi
+3H\left(2V-\dot\phi^2\right)\dot\phi
+\left(\dot\phi^2+2V\right)V'=0.
\end{equation}

During slow roll, $\dot\phi^2\ll 2V$, the first Friedmann equation becomes
\begin{equation}
3H^2\simeq\frac{4V^2+\beta}{2|\alpha|}.
\end{equation}
The equivalence theorem clarifies the stability properties of this model. In the present case,
$g(\mathcal{L}_m)=4\mathcal{L}_m^2+\beta$.
For $\alpha>0$, the effective scalar Lagrangian is
\begin{equation}
P=2(X-V)^2+\frac{\beta}{2},
\end{equation}
for which
\begin{equation}
P_X+2XP_{XX}=4(3X-V)<0
\end{equation}
in the slow-roll regime. This branch therefore suffers from a scalar ghost instability. We hence
focus on $\alpha<0$. After reversing the overall sign of the action, the effective scalar Lagrangian
and its relevant derivatives are
\begin{equation}\label{eq:nlP}
\begin{aligned}
P&=-2(X-V)^2-\frac{\beta}{2},\\
P_X&=4(V-X),\\
P_X+2XP_{XX}&=4(V-3X),\\
c_s^2&=\frac{V-X}{V-3X},
\end{aligned}
\end{equation}
where $X\equiv\dot\phi^2/2$. For $X>0$, the scalar sector is free of ghosts and gradient instabilities
provided
\begin{equation}
V>3X.
\end{equation}
In particular, the sound speed is superluminal, $c_s^2>1$, throughout this healthy domain. Such
superluminal propagation on a non-trivial background does not, by itself, establish the existence of
causal paradoxes \citep{babichev2008kessence}. Nevertheless, it is a feature that should be kept
in mind when assessing the ultraviolet consistency of the effective description.

The sign difference between the linear and nonlinear models is therefore associated with the stability
of their respective scalar kinetic sectors, rather than with a change in the gravitational degrees of
freedom. The nonlinear model has a healthy scalar kinetic sector only on the branch $\alpha<0$ and within the domain $V>3X$.

For $\beta=0$, the exact background equations yield
\begin{equation}
\epsilon_1\equiv-\frac{\dot H}{H^2}
=\frac{6X}{V+3X}.
\end{equation}
Consequently, $\epsilon_1=1$ is reached precisely at $V=3X$. This is also the boundary at which the
no-ghost condition is saturated and the coefficient of $\ddot\phi$ in Eq.~\eqref{eq:nlkg} vanishes.
The background evolution therefore reaches a degenerate point at the end of inflation, beyond
which the present form of the scalar equation does not provide a regular continuation.

\begin{figure*}
\centering
\includegraphics[width=\linewidth]{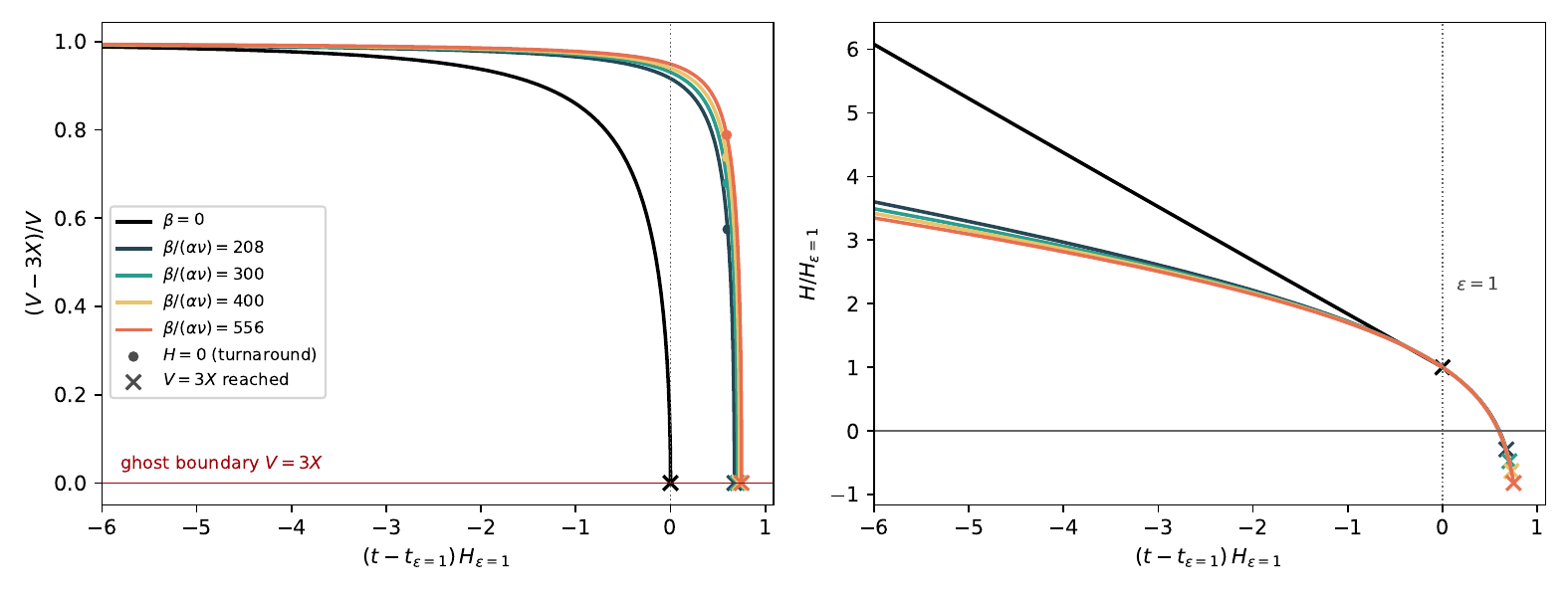}
\caption{End of inflation in the nonlinear model with
$V=\nu\phi^2$, $\alpha=-5\times10^{-2}$, and
$\nu=10^{-3}$, for $\beta=0$ and four negative values
of $\beta/(\alpha\nu)$ spanning the Planck-allowed
interval in Table~\ref{tab:intervals}. The solutions
are obtained by integrating the exact background
equations, with $H$ evolved dynamically. Left:
$(V-3X)/V$, with the boundary $V=3X$ marking the
saturation of the scalar no-ghost condition. Right:
$H/H_{\epsilon=1}$, where $H_{\epsilon=1}$ denotes
the Hubble parameter at the end of inflation.
Time is measured from $\epsilon_1=1$ in units of
$H_{\epsilon=1}^{-1}$. Dots indicate the points at
which $H=0$, corresponding to the turnaround of
the expanding solution, whereas crosses indicate
the boundary $V=3X$. At this boundary, the
coefficient of $\ddot\phi$ in Eq.~\eqref{eq:nlkg}
vanishes, and the numerical integration is
terminated. For $\beta=0$, the boundary is reached
at $\epsilon_1=1$; for the displayed negative values
of $\beta$, it is reached after the turnaround.}
\label{fig:nlend}
\end{figure*}

For negative values of $\beta$ in the parameter intervals considered in Table~\ref{tab:intervals},
the situation is different. Numerical integration of Eqs.~\eqref{eq:nlfried}--\eqref{eq:nlkg} shows
that inflation ends within the healthy domain, at
\begin{equation}
\frac{V-3X}{V}\simeq0.92\text{--}0.95,
\end{equation}
whereas $H^2$ reaches zero after approximately $\Delta N\simeq0.36$ $e$-folds. At $H=0$, the
expanding solution reaches a turnaround. The trajectory subsequently approaches the boundary
$V=3X$, where the coefficient of $\ddot\phi$ in Eq.~\eqref{eq:nlkg} vanishes. The numerical integration is therefore terminated at this
degenerate boundary (Fig.~\ref{fig:nlend}).

Although the sound speed is superluminal in the healthy domain, the sign of the $(\partial\phi)^4$
operator also deserves attention. Expanding the effective Lagrangian gives
\begin{equation}
P=-2X^2+4VX-2V^2-\frac{\beta}{2}.
\end{equation}
The negative coefficient of $X^2$ is opposite to the sign commonly required by forward-limit
positivity bounds for a standard Lorentz-invariant vacuum with conventional asymptotic states
\citep{adams2006causality}. However, the applicability of these bounds to the present model requires care.
In particular, for $V=\nu\phi^2$, one has $P_X=4V=0$ at the potential minimum. The standard
vacuum assumptions underlying the positivity argument are therefore not automatically satisfied
there. We consequently do not draw a definitive conclusion about the existence or absence of a
consistent ultraviolet completion.

During slow roll, the sound speed is close to unity:
\begin{equation}
c_s^2
=\frac{V-X}{V-3X}
\simeq 1+\frac{2X}{V}
=1+\mathcal{O}(\epsilon_1).
\end{equation}
It follows that $s\equiv\dot c_s/(Hc_s)$ is of second slow-roll order, and hence
$r=16c_s\epsilon_1\simeq16\epsilon_1$ at leading order.

Expanding Eq.~\eqref{eq:nlP} to first order in $X$, we obtain
\begin{equation}
P\simeq4V(\phi)X-2V^2-\frac{\beta}{2}.
\end{equation}
The field is not canonically normalized in this approximation. Introducing the canonical field $\varphi$ through
\begin{equation}
\frac{\mathrm{d}\varphi}{\mathrm{d}\phi}
=2\sqrt{V(\phi)},
\end{equation}
the effective potential becomes
\begin{equation}
U(\varphi)=2V(\phi(\varphi))^2+\frac{\beta}{2}.
\end{equation}
The corresponding leading-order slow-roll parameters can be written directly in terms of the original field as
\begin{align}
\epsilon_1
&\simeq
\frac{8|\alpha|VV'^2}{(4V^2+\beta)^2},
\\
\eta
&\simeq
|\alpha|
\left[
\frac{2V''}{4V^2+\beta}
+\frac{V'^2}{V(4V^2+\beta)}
\right],
\label{eq:nlsr}
\\
N
&\simeq
\frac{1}{2|\alpha|}
\int_{\phi_{\rm end}}^{\phi}
\frac{4V^2+\beta}{V'}\,\mathrm{d}\phi.
\label{eq:nlN}
\end{align}
Here, $\eta$ is the potential slow-roll parameter associated with the canonically normalized field
$\varphi$. These expressions agree with the leading-order slow-roll expansion of the exact
background equations. In Eq.~\eqref{eq:nlN}, the positive coefficient $1/(2|\alpha|)$ ensures $N>0$ for the usual rolling direction with
$\phi>\phi_{\rm end}$ and $V'>0$.

\subsection{Quadratic potential: a single parameter}
We now consider the quadratic potential
\begin{equation}
V(\phi)=\nu\phi^2,
\qquad \nu>0.
\end{equation}
For this potential, both the linear and nonlinear models can be expressed, at leading order in slow
roll, in terms of a quadratic effective potential with a constant offset.

In the linear model, the canonically normalized field is
\begin{equation}
\hat\psi=\frac{\phi}{\sqrt{\alpha}},
\end{equation}
where we work on the standard gravitational branch $\alpha>0$ and use reduced Planck units. The corresponding effective potential is
\begin{equation}
U(\hat\psi)
=\alpha\nu\left(\hat\psi^2+c_{\rm L}\right),
\qquad
c_{\rm L}=-\frac{\beta}{2\alpha\nu}.
\label{eq:Ulinear}
\end{equation}

For the nonlinear model, the slow-roll approximation to the scalar Lagrangian is
\begin{equation}
P\simeq 4V(\phi)X-2V(\phi)^2-\frac{\beta}{2}.
\end{equation}
The canonically normalized field $\chi$ is therefore defined by
\begin{equation}
\frac{\mathrm{d}\chi}{\mathrm{d}\phi}
=2\sqrt{V(\phi)}.
\end{equation}
For the branch $\phi>0$, this gives
\begin{equation}
\chi=\sqrt{\nu}\phi^2.
\end{equation}
The effective potential in terms of this field is
\begin{equation}
U(\chi)=2\nu\chi^2+\frac{\beta}{2}.
\end{equation}
Introducing $\hat\chi=\chi/\sqrt{|\alpha|}$, we obtain
\begin{equation}
U(\hat\chi)
=2|\alpha|\nu\left(\hat\chi^2+c_{\rm NL}\right),
\qquad
c_{\rm NL}=\frac{\beta}{4|\alpha|\nu}
=-\frac{\beta}{4\alpha\nu},
\label{eq:Unonlinear}
\end{equation}
where the last equality applies to the branch
$\alpha<0$.

The overall amplitude of each effective potential does not affect the leading-order expressions for
the scalar spectral index and tensor-to-scalar ratio. Consequently, within the quadratic-potential
slow-roll approximation, the dependence of these observables on $(\alpha,\beta,\nu)$ enters through the dimensionless parameter
\begin{equation}
\label{eq:xdef}
x\equiv\frac{\beta}{\alpha\nu}.
\end{equation}
In terms of this parameter,
\begin{equation}
c_{\rm L}=-\frac{x}{2},
\qquad
c_{\rm NL}=-\frac{x}{4}.
\end{equation}
Thus, the two models generate identical families of leading-order $(n_s,r)$ predictions under the parameter mapping
\begin{equation}
x_{\rm NL}=2x_{\rm L}.
\end{equation}
This correspondence concerns their slow-roll inflationary observables for the quadratic potential;
it does not imply equivalence of their full actions or exact background dynamics. Closed-form expressions for the relevant quantities in terms
of the Lambert $W$ function are given in Appendix~\ref{app:lambert}. The appropriate branch,
$W_{-1}$, must be selected to retain the physical solutions satisfying $H^2>0$.

The resulting family of effective potentials has the form
\begin{equation}
U(\varphi)\propto \varphi^2+c,
\end{equation}
where $\varphi$ denotes the corresponding canonically normalized field and $c$ is the dimensionless offset. For $c=0$, standard
quadratic inflation is recovered. For $c<0$, the potential has a zero at $\varphi^2=|c|$, and the inflationary trajectory approaches this zero as
the offset becomes increasingly negative. In the limit $|c|\gg N$, the last $N$ $e$-folds take place on an approximately linear part of the potential.
In this limit, the tensor-to-scalar ratio approaches
\begin{equation}
r\longrightarrow
\frac{4}{N+\tfrac{1}{4}},
\end{equation}
where the constant $1/4$ follows from retaining the end-of-inflation contribution obtained from
$\epsilon=1$. In the opposite limit, $c=0$, the leading-order result is
\begin{equation}
r=\frac{8}{N+\tfrac{1}{2}}.
\end{equation}

Our numerical evaluation shows that $r$ decreases monotonically between these two limits. Hence,
within the considered quadratic-potential family and at leading order in slow roll, we obtain
\begin{equation}
\label{eq:rbound}
\frac{4}{N+\tfrac{1}{4}}
<r\leq
\frac{8}{N+\tfrac{1}{2}},
\qquad c\leq0.
\end{equation}
For $c>0$, the tensor-to-scalar ratio is larger than its value in standard quadratic inflation. In the linear model, $c_{\rm L}>0$ corresponds
to $\beta<0$, whereas in the nonlinear model $c_{\rm NL}>0$ corresponds to $\beta>0$. For $N\leq60$, the lower bound in
Eq.~\eqref{eq:rbound} implies
\begin{equation}
r>\frac{4}{60+\tfrac{1}{4}}
\simeq 0.0664.
\end{equation}
This value is substantially above the BICEP/Keck upper bound. Therefore, within this quadratic-potential family, the parameter region
with $c\leq0$ cannot satisfy the tensor constraint.

\begin{figure*}
\centering
\includegraphics[width=\linewidth]{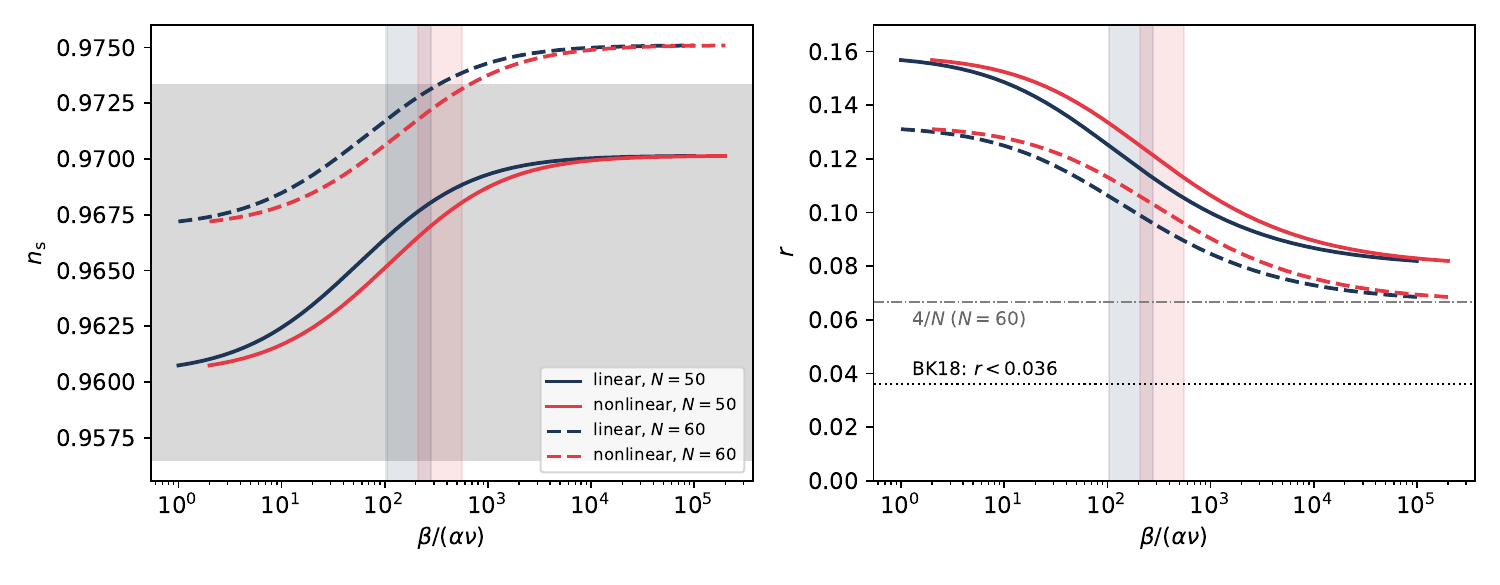}
\caption{Scalar spectral index (left) and
tensor-to-scalar ratio (right) as functions of
$x=\beta/(\alpha\nu)$ for the linear (blue) and
nonlinear (red) models with $V=\nu\phi^2$.
Solid and dashed curves correspond to $N=50$
and $N=60$, respectively. The vertical bands
indicate the parameter intervals for which the
entire trajectory with $N\in[50,60]$ lies within
the Planck TT,TE,EE+\allowbreak lowE+\allowbreak lensing region, as
reported in Table~\ref{tab:intervals}. The grey
shaded region indicates the Planck 2018
constraint $\ns=0.9649\pm0.0042$ at the
$2\sigma$ level. The horizontal lines indicate
the BICEP/Keck upper bound $r<0.036$ and the
asymptotic value $r=4/N$ for $N=60$.}
\label{fig:x}
\end{figure*}

\subsection{Comparison with observations}
\begin{figure}
\centering
\includegraphics[width=\linewidth]{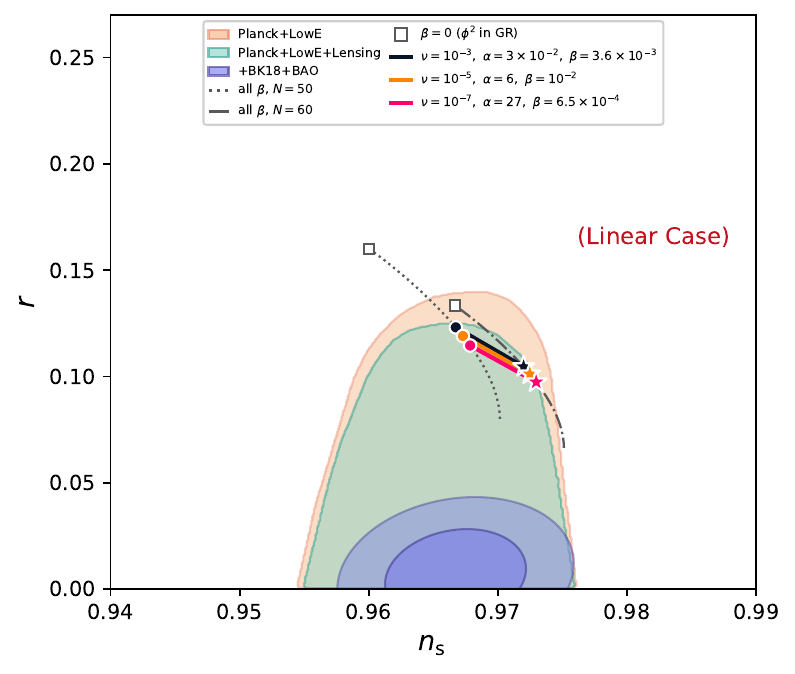}
\caption{Predictions of the linear model with
$V=\nu\phi^2$ in the $(n_s,r)$ plane for
$N\in[50,60]$. Circles and stars correspond to
$N=50$ and $N=60$, respectively. The displayed
parameter values, $x=120$, $166.7$, and $240.7$,
lie within the Planck interval of
Table~\ref{tab:intervals}. Grey dotted and
dash-dotted curves show the predictions obtained
by varying $\beta$ at fixed $N=50$ and $N=60$,
respectively. Open squares correspond to
$\beta=0$, recovering standard quadratic inflation.
The shaded regions show the Planck 2018 and
BICEP/Keck 2018 constraints
\citep{akrami2020planck,2021PhRvL.127o1301A}.}
\label{fig:lin}
\end{figure}

\begin{figure}
\centering
\includegraphics[width=\linewidth]{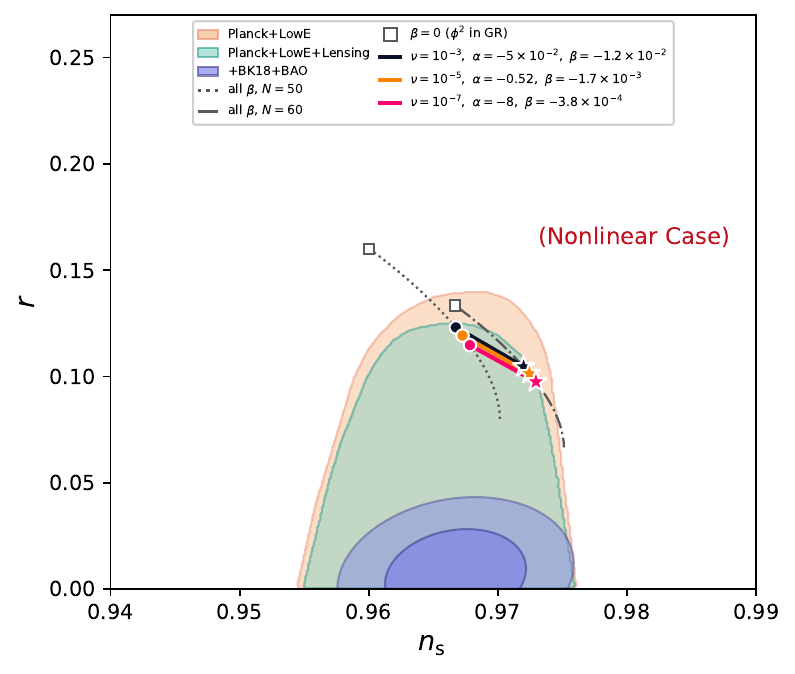}
\caption{As in Fig.~\ref{fig:lin}, but for the
nonlinear model with $x=240$, $326.9$, and $475$.
At leading order in slow roll, the curves coincide
with those of the linear model after applying the
parameter mapping $x_{\rm L}=x_{\rm NL}/2$.}
\label{fig:nl}
\end{figure}

Figure~\ref{fig:x} shows the dependence of $n_s$ and $r$ on the dimensionless parameter $x$. Figures~\ref{fig:lin} and \ref{fig:nl} display
the corresponding trajectories in the $(n_s,r)$ plane, together with the observational constraints from Planck 2018 \citep{akrami2020planck} and
BICEP/Keck 2018 \citep{2021PhRvL.127o1301A}.

We determine the parameter intervals by requiring the entire trajectory over $N\in[50,60]$ to lie within the relevant Planck confidence regions.
The resulting intervals are listed in Table~\ref{tab:intervals}. Within the Planck TT,TE,EE+\allowbreak lowE+\allowbreak lensing interval, the models predict
\begin{equation}
0.9665\lesssim n_s\lesssim0.9731,
\qquad
0.096\lesssim r\lesssim0.125.
\end{equation}
Although these intervals can accommodate the Planck constraints on the scalar spectral index, the predicted tensor-to-scalar ratio is too large
to satisfy the BICEP/Keck upper bound. Thus, no value of $x$ in the considered parameter ranges provides simultaneous compatibility with the
combined constraints.

The ACT DR6 analysis, which favors a somewhat larger scalar spectral index, $n_s=0.974\pm0.003$ \citep{louis2025act},
does not alter this conclusion. The incompatibility is primarily driven by the predicted tensor-to-scalar ratio rather than by the value of $n_s$.

\begin{table*}
\centering
\caption{Indicative intervals of
$x=\beta/(\alpha\nu)$ for which the entire trajectory
with $N\in[50,60]$, sampled in steps of $0.5$,
lies inside the indicated regions of the published
Planck 2018 and BICEP/Keck 2018 constraint plots.
These intervals are obtained by graphical comparison
and are not the result of a dedicated likelihood
analysis.}
\label{tab:intervals}
\begin{ruledtabular}
\begin{tabular}{lcc}
Data & Linear ($\alpha,\beta>0$)
& Nonlinear ($\alpha,\beta<0$)\\
\hline
Planck TT,TE,EE+lowE
& $40\lesssim x\lesssim1.49\times10^3$
& $80\lesssim x\lesssim2.98\times10^3$\\
\quad + lensing
& $104\lesssim x\lesssim278$
& $208\lesssim x\lesssim556$\\
\quad + lensing + BK18 + BAO
& none
& none\\
\end{tabular}
\end{ruledtabular}
\end{table*}

\subsection{The price of the offset}
The improvement in the scalar spectral-index predictions relative to standard quadratic inflation is associated with a negative constant offset
$c<0$ in the effective potential. This offset also has consequences for the post-inflationary evolution.

In the linear model, the effective potential is $U=V-\beta/2$. For $\alpha>0$ and $\beta>0$, its minimum contains a negative vacuum-energy contribution,
\begin{equation}
\rho_{\rm vac}^{\rm (L)}
=-\frac{\beta}{2},
\qquad
\Lambda_{\rm eff}
=-\frac{\beta}{2\alpha}<0.
\end{equation}
In the nonlinear model, the effective potential obtained in the slow-roll canonical description is $U=2V^2+\beta/2$. For $\beta<0$, its minimum
likewise contains a negative constant contribution,
\begin{equation}
\rho_{\rm vac}^{\rm (NL)}
=\frac{\beta}{2}<0.
\end{equation}
The two models therefore share a negative-offset feature in their effective inflationary potentials, although their exact post-inflationary dynamics
need not be the same.

For the linear model, suppose that reheating produces a standard relativistic plasma minimally coupled to gravity and that the scalar field has
settled at the minimum of its potential. The subsequent Friedmann equation then takes the form
\begin{equation}
3M_{\rm Pl}^2H^2
=\rho_r(a)-|\rho_{\rm vac}^{\rm (L)}|,
\qquad
\rho_r(a)\propto a^{-4}.
\end{equation}
Under these assumptions, the expanding solution
reaches $H=0$ at
\begin{equation}
\frac{a_{\rm turn}}{a_{\rm end}}
=
\left(
\frac{\rho_{r,\rm end}}
{|\rho_{\rm vac}^{\rm (L)}|}
\right)^{1/4}.
\label{eq:aturn}
\end{equation}
If the negative vacuum contribution is a substantial fraction of the energy density at the end of inflation, Eq.~\eqref{eq:aturn} implies that the
turnaround occurs after a relatively small increase in the scale factor. In particular, for the parameter choices for which
$a_{\rm turn}/a_{\rm end}\simeq1.3$--$1.6$,
the corresponding interval is
\begin{equation}
\Delta N
=\ln\left(\frac{a_{\rm turn}}{a_{\rm end}}\right)
\simeq0.26\text{--}0.47.
\end{equation}
Thus, under the stated reheating assumptions, the negative vacuum contribution in the linear model prevents a prolonged standard
radiation-dominated expansion and leads to a  turnaround.

For the nonlinear model, the same conclusion cannot be obtained by directly applying the standard radiation Friedmann equation. The
nonlinear dependence of the action on $\mathcal{L}_m$ modifies the matter sector, and its effects during reheating and the subsequent
radiation-dominated phase must be included. Consequently, establishing a turnaround or recollapse in this model requires a separate
post-inflationary analysis.

The negative offset therefore represents a potential post-inflationary viability problem, particularly for the linear model under standard
reheating assumptions. A complete assessment requires specifying the reheating mechanism and following the evolution of the matter sector
consistently with the underlying action. We therefore do not infer a universal recollapse of both models, nor do we claim that an
independent positive vacuum contribution is necessarily required, without such an analysis.

\section{Coupled models: ghost instability}\label{sec:coupled}
\subsection{Model and background}
Any departure from the separable class \eqref{eq:sepclass} makes $f_Q$ depend on $\mathcal{L}_m$, on $Q$, or on both. Since the
$Q$-dependence alone reproduces the known pathologies of $f(Q)$ gravity \citep{gomes2024pathological,jimenez2026nonviability},
we focus on the coupling that is specific to $f(Q,\mathcal{L}_m)$, taking as a representative example
\begin{equation}\label{eq:coupled}
f=-\alpha Q+2\mathcal{L}_m+\gamma\,Q\,\mathcal{L}_m,
\end{equation}
together with its two pieces, $\gamma QX$ and $-\gamma QV$, obtained by replacing $\mathcal{L}_m$ in the last term with $X$ or $-V$. We set $\beta=0$ to avoid the vacuum-energy problem of Sec.~\ref{sec:separable}. Now $f_Q=-\alpha+\gamma\mathcal{L}_m$ varies along the background.
We define the effective Planck mass $M_T^2\equiv-f_Q=\alpha-\gamma\bar{\mathcal{L}}_m$, which determines the tensor normalization (Sec.~\ref{sec:pert}).

Varying the minisuperspace action, or equivalently using Eqs.~\eqref{eq:F1} and \eqref{eq:F2}, gives
\begin{align}
&3H^2\Big(\alpha+\gamma V-\tfrac32\gamma\dot\phi^2\Big)
=\frac{\dot\phi^2}{2}+V,
\label{eq:cF1}\\
&M_T^2\left(2\dot H+3H^2\right)
=V-\frac{\dot\phi^2}{2}
+2\gamma H\dot\phi\left(\ddot\phi-V'\right),
\label{eq:cF2}\\
&\left(1+3\gamma H^2\right)\left(\ddot\phi+V'\right)
+3H\dot\phi\left(1+3\gamma H^2+2\gamma\dot H\right)=0,
\label{eq:ckg}
\end{align}
where
\[
M_T^2=\alpha+\gamma V-\frac{\gamma}{2}\dot\phi^2.
\]
The last term in Eq.~\eqref{eq:cF2} is the contribution proportional to $\dot f_Q$ in Eq.~\eqref{eq:F2}, where
\[
\dot f_Q=\gamma\dot\phi(\ddot\phi-V').
\]
In the slow-roll regime, Eqs.~\eqref{eq:cF1} and \eqref{eq:ckg} give, at leading order,
\begin{equation}\label{eq:csr}
H^2\simeq\frac{V}{3(\alpha+\gamma V)},\qquad
\dot\phi\simeq-\frac{V'}{3H},\qquad
\epsilon\simeq\frac{\alpha}{2}
\left(\frac{V'}{V}\right)^2,
\end{equation}
and
\[
N\simeq\int
\frac{V}{(\alpha+\gamma V)V'}\,d\phi.
\]
For $V=\nu\phi^2$ and $\gamma>0$, the slow-roll expression for $H^2$ approaches the constant $1/(3\gamma)$ at large
field values. Thus, the coupling drives the background expansion towards a quasi-de Sitter regime. This background
property alone, however, does not establish the stability of the scalar perturbations.

\subsection{What a background-only analysis would conclude}
\begin{figure}
\centering
\includegraphics[width=\linewidth]{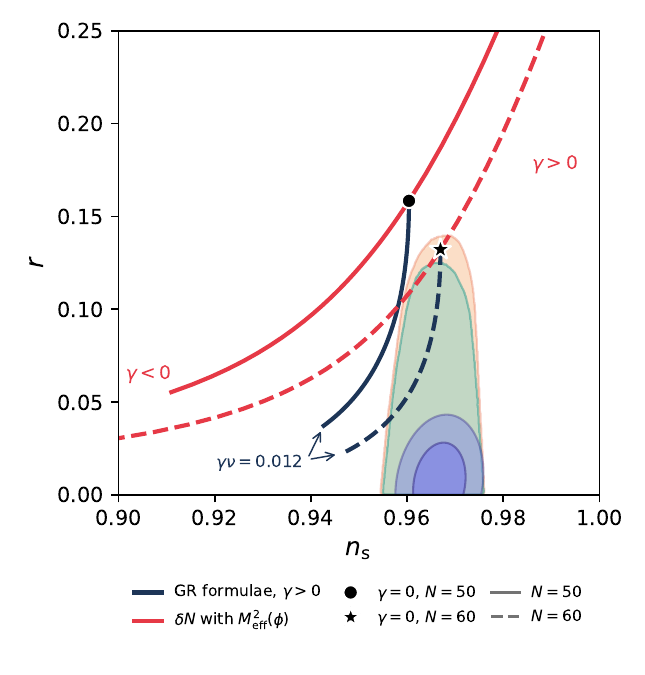}
\caption{Background-level estimates for the coupled model
\eqref{eq:coupled} with $V=\nu\phi^2$. Blue: the GR formulae
\eqref{eq:obsGR} applied to the modified background,
$0\leq\gamma\nu\leq0.012$. Red: the heuristic estimate
including the field-dependent effective Planck mass,
$-0.0022\leq\gamma\nu\leq0.0034$. Solid: $N=50$; dashed:
$N=60$. The filled circle and star mark $\gamma=0$, i.e.,
standard quadratic inflation in GR, at $N=50$ and $60$,
respectively. The two prescriptions move in opposite
directions. Neither curve represents a physical prediction:
Sec.~\ref{sec:pert} shows that the model propagates a ghost
at regular background points for every $\gamma\neq0$.
The range $\gamma\nu\leq0.012$ shown here lies below the
coupling, $\gamma\nu\simeq0.014$, at which the degenerate
point discussed in Sec.~\ref{sec:pert} is crossed within
the last 60 $e$-folds.}
\label{fig:bg}
\end{figure}

Inserting the Hubble-flow parameters of this background into the GR formulae, Eq.~\eqref{eq:obsGR}, gives, in units
$\alpha=1$,
\[
n_s\simeq1-4\epsilon-4\gamma\nu,
\qquad
r=16\epsilon,
\]
with $\epsilon$ exponentially suppressed in $\gamma\nu N$. Positive $\gamma$ then lowers $r$ towards the BICEP/Keck
region; for example, $\gamma\nu=0.008$ gives $(n_s,r)=(0.957,0.043)$ at $N=60$ (blue curves in Fig.~\ref{fig:bg}). Taken at face value, this background-only prescription would suggest that the coupling improves the observational compatibility of quadratic inflation.

The GR formulae are, however, not justified when the effective Planck mass varies and the scalar perturbations are coupled
to the metric sector. One may construct an alternative illustrative estimate by combining the $\delta N$ expression
with the decoupled-inflaton normalization $1+3\gamma H^2$ and the tensor normalization $M_T^2$. This gives the heuristic expression
\[
r_{\rm heur}\simeq
16\epsilon(1+3\gamma H^2)\frac{M_T^2}{\alpha}.
\]
This expression is not a controlled prediction for the tensor-to-scalar ratio: obtaining the latter requires the
reduced scalar quadratic action and the properly normalized propagating scalar modes. Nevertheless, as an illustrative
background-level diagnostic, it moves in the opposite direction: positive $\gamma$ raises the estimated $r$,
whereas negative $\gamma$, close to the point at which $1+3\gamma H^2$ vanishes, lowers it (red curves in
Fig.~\ref{fig:bg}). The disagreement between these two prescriptions illustrates why background evolution alone
cannot determine the observational predictions of the model. The scalar perturbations must instead be derived from the
full theory, as we do next.

\subsection{Quadratic action for perturbations}\label{sec:pert}
\paragraph*{Gauge fixing:}
In the covariant formulation, the flat, torsion-free connection is
\[
\Gamma^\alpha{}_{\mu\nu}
=\frac{\partial x^\alpha}{\partial\xi^\lambda}
\,\partial_\mu\partial_\nu\xi^\lambda,
\]
where $\xi^\lambda$ are four Stückelberg fields \citep{jimenez2018coincident}. The action
$S[g,\xi,\phi]$ is diffeomorphism invariant. Under $x^\mu\rightarrow x^\mu+\epsilon^\mu$, the fields transform
as
\begin{gather*}
\delta g_{\mu\nu}=-2\nabla_{(\mu}\epsilon_{\nu)},\\
\delta\xi^\lambda=-\epsilon^\mu\partial_\mu\xi^\lambda,
\qquad
\delta\phi=-\epsilon^\mu\partial_\mu\phi.
\end{gather*}
Invariance under arbitrary $\epsilon^\nu$ yields the off-shell Noether identity
\begin{equation}\label{eq:noether}
2\nabla_\mu\mathcal{E}^{\mu}{}_{\nu}
-\mathcal{E}_\lambda\,\partial_\nu\xi^\lambda
-\mathcal{E}_\phi\,\partial_\nu\phi=0,
\end{equation}
where $\mathcal{E}^{\mu\nu}$, $\mathcal{E}_\lambda$ and $\mathcal{E}_\phi$ denote the variational derivatives with respect to $g_{\mu\nu}$, $\xi^\lambda$ and $\phi$, respectively, with the corresponding index conventions understood. When the metric and inflaton equations hold,
Eq.~\eqref{eq:noether} implies $\mathcal{E}_\lambda\partial_\nu\xi^\lambda=0$. Since $\partial_\nu\xi^\lambda$ is invertible, the
connection equation $\mathcal{E}_\lambda=0$ then follows.

We impose the coincident gauge, $\xi^\lambda=x^\lambda$, which completely fixes the local diffeomorphism freedom. Indeed, preserving this condition
requires $\delta\xi^\lambda=0$, and hence $\epsilon^\lambda=0$. Imposing this gauge directly in the
action is legitimate provided that all metric perturbations are retained \citep{motohashi2016fundamental}. Our analysis
is restricted to perturbations around the trivial cosmological branch, in which the background Stückelberg
fields coincide with the Cartesian coordinates and $Q=6H^2$. This is the branch of symmetric teleparallel gravity considered here and the setting in which the background relation $Q=6H^2$ holds \citep{gomes2024pathological}.

In the coincident gauge, the connection is fixed and no local diffeomorphism freedom remains. Consequently, all scalar metric perturbations must be retained. We write
\begin{align}
ds^2={}&-(1+2\Phi)dt^2
+2a\,\partial_iB\,dt\,dx^i\nonumber\\
&+a^2\left[(1-2\psi)\delta_{ij}
+2\partial_i\partial_jE\right]dx^idx^j,
\label{eq:metricpert}
\end{align}
and
\[
\phi=\bar\phi(t)+\delta\phi.
\]
We expand $\frac12\int d^4x\sqrt{-g}\,f$ to second order in perturbations, with $Q$ computed from Eq.~\eqref{eq:Qdef}
using $Q_{\alpha\mu\nu}=\partial_\alpha g_{\mu\nu}$ in the coincident gauge, and then transform to Fourier space.
We use the background equations \eqref{eq:cF1}--\eqref{eq:ckg} and integrate by parts. The calculation was carried out symbolically; the procedure
and consistency checks are described in Appendix~\ref{app:pert}. In particular, the zeroth-order action reproduces the minisuperspace Lagrangian, and all first-order terms vanish when the background equations are imposed.

\paragraph*{Tensor sector:}
For $\delta g_{ij}=a^2h_{ij}$, with $h_{ij}$ transverse and traceless, we find
\begin{equation}\label{eq:tensor}
S_T^{(2)}
=\frac18\int dt\,d^3x\;a^3M_T^2
\left[\dot h_{ij}^2-a^{-2}(\partial_kh_{ij})^2\right],
\end{equation}
after the would-be mass term cancels on the background. Tensor modes therefore propagate at the speed of light,
and they are free of tensor ghosts when $M_T^2>0$. This condition is satisfied throughout the domain of Theorem~2 below. To see this, we use the exact background identity
\begin{equation}\label{eq:sid}
\begin{aligned}
1+3\gamma H^2
&=\frac{2M_T^2-\alpha}{D_F},\\
D_F&\equiv
\alpha+\gamma V-\frac32\gamma\dot\phi^2
=\frac{\rho}{3H^2},
\end{aligned}
\end{equation}
where
$\rho=\dot\phi^2/2+V>0$. Thus, for $\alpha>0$,
$1+3\gamma H^2>0$ implies $M_T^2>\alpha/2>0$.

\paragraph*{Scalar sector:}
For the scalar perturbations in Eq.~\eqref{eq:metricpert}, the terms linear in $\dot\Phi$ and $\dot B$ are integrated
by parts. The fields $\Phi$ and $B$ then appear without time derivatives, and $B$ appears only linearly ($M_{BB}=0$). Their Hessian has determinant $-M_{\Phi B}^2$, where
\begin{equation}\label{eq:auxhessian}
M_{\Phi B}
=\frac{a^2k^2\,\mathcal{D}}
{4(3\gamma H^2-1)\,\mathcal{B}}.
\end{equation}
For $k\neq0$, $\mathcal{D}\neq0$, $\mathcal{B}\neq0$ and $3\gamma H^2\neq1$, this determinant is nonzero, so both auxiliary fields can be
eliminated through their algebraic equations. The remaining Lagrangian has the form
\begin{equation}\label{eq:scalarL}
L^{(2)}
=\frac12\dot Y^{\rm T}\mathcal{K}\dot Y
+\dot Y^{\rm T}\mathcal{N}Y
-\frac12Y^{\rm T}\mathcal{W}Y,
\end{equation}
with $Y=(\psi,E,\delta\phi)$.

For $\gamma=0$, the kinetic matrix $\mathcal{K}$ has rank one, as expected for the scalar sector of the corresponding
GR model after the lapse and shift constraints have been eliminated. For $\gamma\neq0$, we find exactly
\begin{align}
\mathcal{K}_{EE}&=0,
\label{eq:KEE}\\
\mathcal{K}_{\psi E}
&=-\frac{\gamma\dot\phi\,\mathcal{A}\,\mathcal{T}}
{\mathcal{D}},
\nonumber\\
\mathcal{K}_{E\delta\phi}
&=-\frac{\gamma H\dot\phi\,\mathcal{U}}
{\mathcal{D}}
\qquad (a=k=1),
\label{eq:Koff}\\
\det\mathcal{K}
&=-\frac{a^9k^4\gamma^2\dot\phi^2
(1+3\gamma H^2)^2
\mathcal{A}\,\mathcal{B}\,\mathcal{S}^2}
{2\,\mathcal{D}^2}.
\label{eq:detQL}
\end{align}
Here,
\begin{align}
\mathcal{A}
&=\gamma\dot\phi^2(1+3\gamma H^2)-\alpha,
\\
\mathcal{B}
&=\gamma\dot\phi^2(1+27\gamma^2H^4)
-\alpha(1+3\gamma H^2),
\\
\mathcal{S}
&=-3H\dot\phi\,(\alpha-2\gamma\dot\phi^2)
+2V'\mathcal{A},
\end{align}
and $\mathcal{D}$, $\mathcal{T}$ and $\mathcal{U}$ are polynomials in $(\alpha,\gamma,H,\dot\phi,V')$ given in
Appendix~\ref{app:pert}. The background equations for $(\dot H,\ddot\phi)$ are regular provided that
$\mathcal{B}\neq0$ and $3\gamma H^2\neq1$; the Jacobian of this system is
\[
-\frac{6\mathcal{B}}{3\gamma H^2-1}.
\]

\textit{Theorem 1 (existence of a ghost).}
Let $\alpha>0$ and $\gamma\neq0$, and consider a regular background point satisfying $\rho>0$, $\mathcal{B}\neq0$,
$3\gamma H^2\neq1$, $\mathcal{D}\neq0$ and $k\neq0$. If $\dot\phi\neq0$ and $(\mathcal{A}\mathcal{T},H\mathcal{U})\neq(0,0)$,
then the kinetic matrix $\mathcal{K}$ has at least one negative and at least one positive eigenvalue.

\textit{Proof.}
The kinetic matrix is real and symmetric. By Eq.~\eqref{eq:KEE}, the principal $2\times2$ submatrix
on $(\psi,E)$ has determinant $-\mathcal{K}_{\psi E}^2$, while the principal submatrix on $(E,\delta\phi)$ has determinant
$-\mathcal{K}_{E\delta\phi}^2$. By the assumptions of the theorem, at least one of these determinants is strictly
negative. The corresponding submatrix therefore has one negative and one positive eigenvalue. Cauchy's interlacing
theorem then implies $\lambda_1<0<\lambda_3$ for the eigenvalues $\lambda_1\leq\lambda_2\leq\lambda_3$ of $\mathcal{K}$. \hfill$\square$

\textit{Theorem 2 (exactly one ghost).}
Under all the hypotheses of Theorem~1, suppose in addition that $\mathcal{S}\neq0$ and either $\gamma>0$, or
$\gamma<0$ with $1+3\gamma H^2>0$. Then $\det\mathcal{K}<0$, and $\mathcal{K}$ has exactly one negative eigenvalue.

\textit{Proof.}
Equation~\eqref{eq:sid} implies
\[
1-3\gamma H^2
=\frac{\alpha-2\gamma\dot\phi^2}{D_F},
\]
with $D_F>0$. Define
\[
y\equiv\frac{\gamma\dot\phi^2}{\alpha},
\qquad
u\equiv3\gamma H^2,
\]
so that $27\gamma^2H^4=3u^2$. The background identity then gives
\[
1-u=\frac{\alpha}{D_F}(1-2y).
\]

For $\gamma>0$, one has $y>0$ and $u>0$. Moreover, $y<\tfrac12$ is equivalent to $u<1$, whereas $y>\tfrac12$ is equivalent to $u>1$. In the first case,
\[
\frac{\mathcal{A}}{\alpha}
=y(1+u)-1
<\frac12(1+u)-1
=\frac{u-1}{2}<0,
\]
and
\begin{align}
\frac{\mathcal{B}}{\alpha}
&=y(1+3u^2)-(1+u)
<\frac12(1+3u^2)-(1+u)\nonumber\\
&=\frac12(3u+1)(u-1)<0.\nonumber
\end{align}
For $y>\tfrac12$ and $u>1$, the inequalities are reversed:
\[
\frac{\mathcal{A}}{\alpha}
>\frac{u-1}{2}>0,
\qquad
\frac{\mathcal{B}}{\alpha}
>\frac12(3u+1)(u-1)>0.
\]
Thus $\mathcal{A}\mathcal{B}>0$ for $\gamma>0$.

For $\gamma<0$, one has $y<0$ and $u<0$. Under the additional condition $1+u>0$,
\[
\frac{\mathcal{A}}{\alpha}
=y(1+u)-1<0,
\qquad
\frac{\mathcal{B}}{\alpha}
=y(1+3u^2)-(1+u)<0.
\]
Hence $\mathcal{A}\mathcal{B}>0$ also for $\gamma<0$ in the stated domain. Since $\mathcal{S}\neq0$ and
$1+u>0$ in the domain of the theorem, Eq.~\eqref{eq:detQL} therefore gives $\det\mathcal{K}<0$. From Theorem~1, $\lambda_1<0<\lambda_3$. Consequently,
\[
\lambda_2
=\frac{\det\mathcal{K}}{\lambda_1\lambda_3}>0,
\]
so there is exactly one negative eigenvalue. \hfill$\square$

\vspace{0.3em}

\noindent\emph{Remark.}
Theorems~1 and 2 are local statements about regular background points satisfying their explicitly stated
conditions. Their application to inflationary solutions below is based on the numerical integration of the background equations for the specified potential and parameter ranges. They do not constitute an exhaustive proof for every possible functional form of
$f(Q,\mathcal{L}_m)$.

All entries of $\mathcal{K}$ scale as $a^3$. Those in the $E$ row and column carry an additional factor $k^2$,
because $E$ enters the metric through $\partial_i\partial_jE$. For $k\neq0$, the change of variable $\tilde E=k^2E$ removes this explicit
wavenumber dependence from the kinetic matrix. Since this is an invertible field redefinition, the inertia (the numbers of positive, negative and zero eigenvalues) is unchanged. The magnitudes of individual eigenvalues, by contrast, depend on the field normalization and are
not invariant under general invertible field redefinitions; their signs and the inertia are the relevant stability information.

The excluded sets correspond to points at which the quadratic action or the background system becomes
degenerate or singular. They must therefore be treated separately. At $\mathcal{D}=0$, the auxiliary sector
cannot be eliminated using the algebraic procedure described above. At $\mathcal{B}=0$, the background equations become singular, and at
$3\gamma H^2=1$ (equivalently $2\gamma\dot\phi^2=\alpha$), both $\mathcal{A}$ and $\mathcal{B}$ vanish. At $\dot\phi=0$, the off-diagonal
entries in Eq.~\eqref{eq:Koff} vanish and the kinetic matrix may lose rank. At $1+3\gamma H^2=0$ (which can occur for $\gamma<0$), the determinant in Eq.~\eqref{eq:detQL} vanishes, although Theorem~1 can still apply if its other hypotheses are satisfied.

\begin{figure*}
\centering
\includegraphics[width=0.8\linewidth]{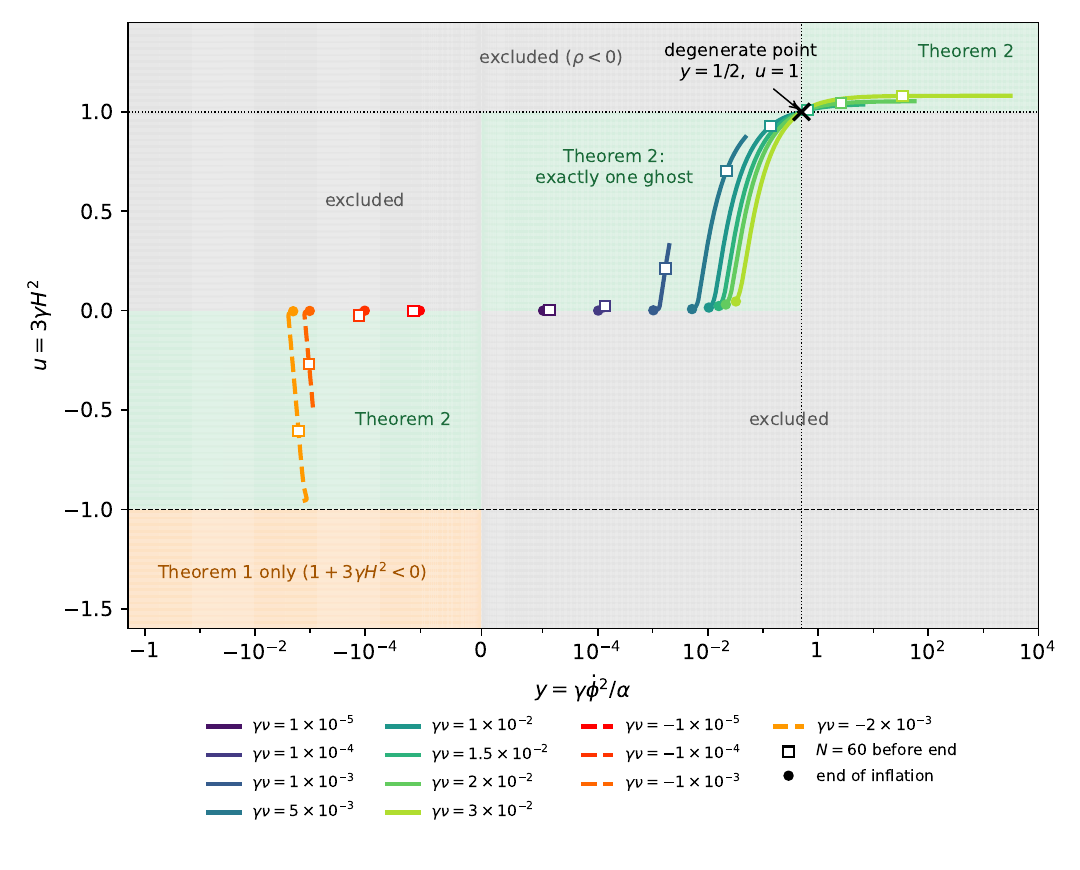}
\caption{Exact inflationary trajectories of the coupled
model \eqref{eq:coupled} with $V=\nu\phi^2$ and
$\alpha=1$ in the plane of
$y=\gamma\dot\phi^2/\alpha$ and $u=3\gamma H^2$,
over the last 100 $e$-folds (solid: $\gamma>0$;
dashed: $\gamma<0$). Green: domain of Theorem~2,
where exactly one ghost is present at regular points;
orange: $\gamma<0$ with $1+3\gamma H^2<0$, where
Theorem~1 may apply but Theorem~2 does not. Grey:
regions excluded by the exact Friedmann equation
($\rho<0$) or by
$\mathrm{sign}(y)\neq\mathrm{sign}(u)$. Squares
mark 60 $e$-folds before the end of inflation, and
dots mark the end of inflation. For
$\gamma\nu\gtrsim10^{-2}$, the trajectories pass
through the degenerate point $y=\tfrac12$, $u=1$
(cross), where the regularity assumptions of the
theorems fail.}
\label{fig:domain}
\end{figure*}

\begin{figure*}
\centering
\includegraphics[width=\linewidth]{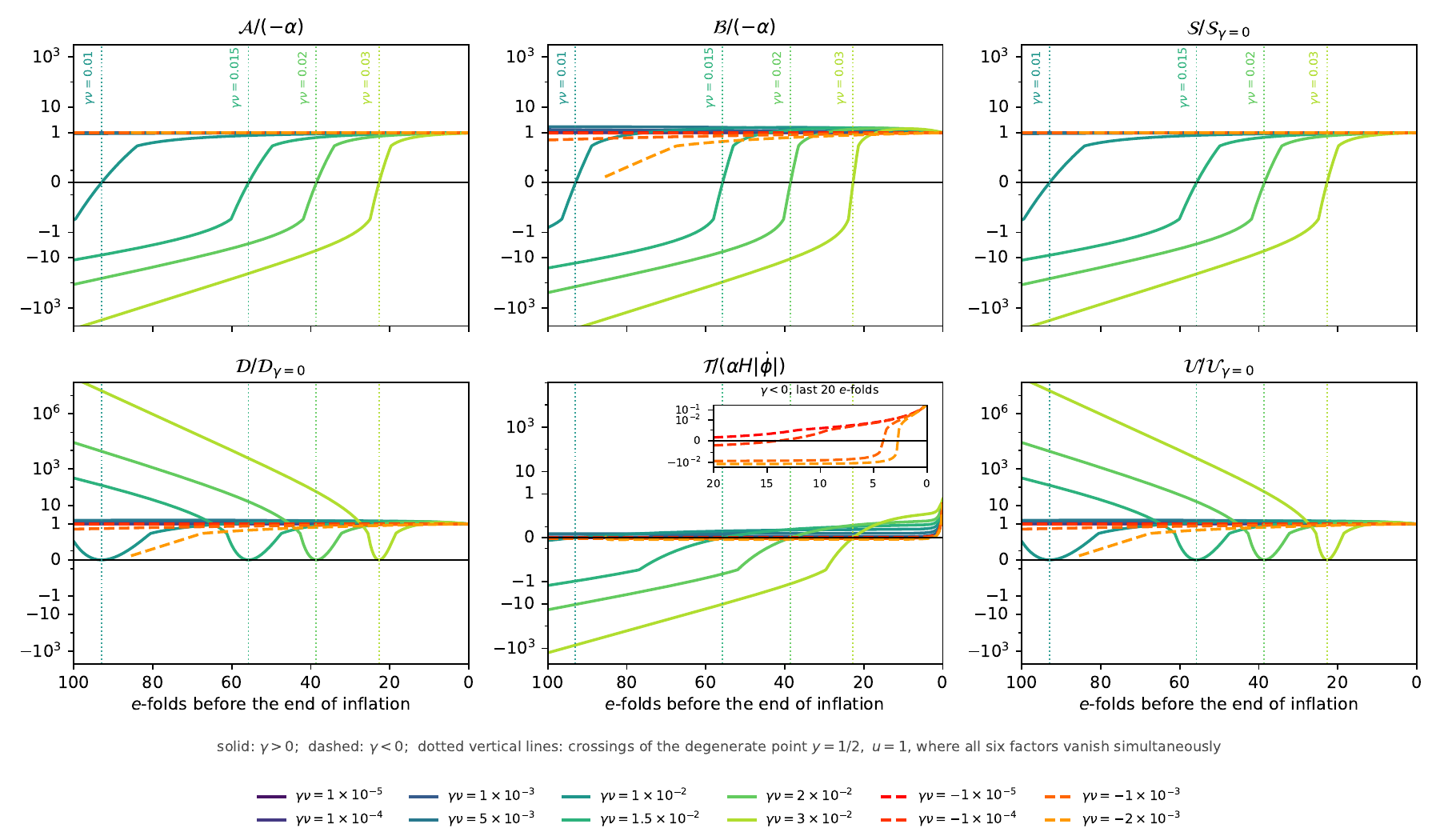}
\caption{The factors entering Theorems~1 and 2,
Eqs.~\eqref{eq:Koff} and \eqref{eq:detQL}, along the
trajectories of Fig.~\ref{fig:domain}, normalized by
their values at $\gamma=0$. The factor $\mathcal{T}$
is instead normalized by $\alpha H|\dot\phi|$, since
its $\gamma=0$ value, $2\alpha\ddot\phi$, is small
during slow roll. Solid: $\gamma>0$; dashed:
$\gamma<0$. Dotted vertical lines, labeled by
$\gamma\nu$, mark the crossings of the degenerate
point $y=\tfrac12$, $u=1$, where all six factors
vanish simultaneously according to the numerical
evaluation. $\mathcal{D}$ and $\mathcal{U}$ are
positive along the regular portions of the plotted
trajectories. The inset shows $\mathcal{T}$ for
$\gamma<0$ over the last 20 $e$-folds.}
\label{fig:factors}
\end{figure*}

To examine these conditions on actual inflationary solutions, we integrated Eqs.~\eqref{eq:cF1}--\eqref{eq:ckg} for $V=\nu\phi^2$ over the last 100 $e$-folds, for $-2\times10^{-3}\leq\gamma\nu\leq3\times10^{-2}$ (Figs.~\ref{fig:domain} and \ref{fig:factors}). For the
parameter choices and initial conditions considered, $\mathcal{D}$ and $\mathcal{U}$ remain positive along
the regular portions of the trajectories, apart from the isolated degenerate points discussed below. Thus, away from those points, the auxiliary sector can be eliminated and the condition of Theorem~1 is satisfied provided the remaining hypotheses of that theorem hold. The factor $\mathcal{T}$ changes sign shortly before the end of inflation for $\gamma<0$. At $\gamma=0$, it reduces to $2\alpha\ddot\phi$; at the numerical
sign-change points, $\mathcal{U}\neq0$, so the condition in Theorem~1 remains satisfied.

For $\gamma\nu\gtrsim10^{-2}$, the trajectories start in the region $y>\tfrac12$, $u>1$ and, as required by Eq.~\eqref{eq:sid}, pass through the degenerate point $y=\tfrac12$, $u=1$. At this point, $\mathcal{A}$, $\mathcal{B}$ and $\mathcal{S}$ vanish
algebraically. The numerical evaluation of the appendix expressions also finds that $\mathcal{T}$, $\mathcal{U}$ and $\mathcal{D}$ vanish
there, so the reduced quadratic action becomes degenerate and the regularity assumptions of the theorems fail. This crossing occurs approximately
93, 56, 39 and 23 $e$-folds before the end of inflation for $\gamma\nu=0.01$, $0.015$, $0.02$ and $0.03$,
respectively. Thus, for $\gamma\nu\gtrsim0.014$, the crossing lies within the last 60 $e$-folds. On either side of the crossing, at regular points satisfying the hypotheses of Theorem~2, there is exactly one negative eigenvalue. The theorem itself does not apply at the
degenerate crossing.

For $\gamma<0$, the condition $1+3\gamma H^2>0$ restricts the available number of $e$-folds. In the numerical solutions considered here,
60 $e$-folds can be obtained only for $|\gamma|\nu\lesssim3\times10^{-3}$.

As an independent numerical check, we computed the eigenvalues of $\mathcal{K}$ using 60-digit arithmetic
at 48 sampled slow-roll background configurations: $10^{-4}\leq\gamma\nu\leq2\times10^{-2}$ and
$-10^{-3}\leq\gamma\nu\leq-10^{-4}$, with $N=50,60$ and $10^{-2}\leq k/(aH)\leq10^2$.
At all these sampled regular points, we find exactly one negative eigenvalue (Fig.~\ref{fig:ghost}). The
same result is obtained at 40 points along each of the regular portions of the exact trajectories in
Fig.~\ref{fig:domain}, evaluated at $k/(aH)=10^{-2}$, $1$ and $10^2$. We also verified Eq.~\eqref{eq:detQL} against a direct numerical Schur
complement at randomly selected regular points, finding a relative agreement of $10^{-46}$.

\paragraph{The two pieces of the coupling:}
We now consider separately $f=-\alpha Q+2\mathcal{L}_m-\gamma QV$, corresponding
to a field-dependent coefficient of $Q$, and $f=-\alpha Q+2\mathcal{L}_m+\gamma QX$. For the first model, the determinant is
\begin{equation}\label{eq:detQV}
\det\mathcal{K}
=-\frac{a^9k^4\gamma^2V'^2\dot\phi^2
(2\alpha-\gamma\dot\phi^2)^2}
{8\left(
3\gamma^2H^2V'\dot\phi-4\alpha H
+2\gamma H\dot\phi^2-\gamma V'\dot\phi
\right)^2}.
\end{equation}
It is strictly negative at regular points where $\gamma V'\dot\phi\neq0$ and the denominator is finite
and nonzero. In particular, this condition is satisfied during ordinary slow-roll evolution with nonzero $\gamma$, $V'$ and $\dot\phi$, provided the denominator does not vanish.

For the second model, in units $\alpha=a=k=1$, we find
\begin{align}
\det\mathcal{K}
={}&\gamma^2\dot\phi^2(\gamma\dot\phi^2-2)
\nonumber\\
&\times
\frac{
\left[2(1+3\gamma H^2)
-\gamma\dot\phi^2(1-9\gamma H^2)\right]
\mathcal{S}_1^2
}{8\,\mathcal{S}_2^2},
\label{eq:detQX}
\end{align}
where $\mathcal{S}_{1,2}$ are polynomials in $(H,\dot\phi,V',\gamma)$ given in Appendix~\ref{app:pert}.
The factor $\gamma\dot\phi^2-2$ is negative whenever the tensor modes are healthy, since $M_T^2=1-\gamma X=1-\gamma\dot\phi^2/2>0$.
For the determinant to be negative, the remaining bracket must be positive. The exact condition is
\[
2(1+3\gamma H^2)>
\gamma\dot\phi^2(1-9\gamma H^2).
\]
Thus, the bracket is positive in the regime where $1+3\gamma H^2$ is positive and bounded away from zero,
and the correction proportional to $\gamma\dot\phi^2$ is sufficiently small. Under this
condition, and at regular points with $\gamma\neq0$, $\dot\phi\neq0$, $\mathcal{S}_1\neq0$
and $\mathcal{S}_2\neq0$, Eq.~\eqref{eq:detQX} gives $\det\mathcal{K}<0$. Hence the kinetic matrix has an
odd number of negative eigenvalues, and in particular at least one ghost. By Sylvester's law of inertia, the
numbers of positive and negative eigenvalues cannot be changed by an invertible linear field redefinition. Both pieces of the coupling therefore exhibit a scalar ghost wherever the corresponding regularity and sign conditions are satisfied.

\paragraph{Behavior as $\gamma\to0$:}
For the background configurations considered in our numerical analysis, the two additional eigenvalues
decrease in magnitude as $\gamma\to0$. In the chosen field basis and normalization, the negative eigenvalue
scales approximately as $\gamma^2$ for moderate couplings and as $\gamma$ for extremely small
couplings; the second additional eigenvalue also vanishes in this limit. These scaling statements refer
to the specified field basis and numerical background configurations, since eigenvalue magnitudes are not
invariant under general field redefinitions. The rank of $\mathcal{K}$ changes from three for nonzero $\gamma$ at the regular points considered to one at
$\gamma=0$, where the corresponding GR constraint structure is recovered. This change in rank indicates
that the limit $\gamma\to0$ is singular at the level of the quadratic kinetic and constraint structure.
However, the quadratic action alone does not determine whether the additional modes decouple smoothly or
become strongly coupled as the coupling approaches zero. Establishing this requires an analysis of the canonically normalized nonlinear interactions, which
is beyond the scope of the present work.

\subsection{Consequences}

The background-based estimates shown in Fig.~\ref{fig:bg} should therefore not be interpreted as physical perturbation predictions of the coupled theory. The presence of a scalar ghost in the coupled scalar--gravity sector signals a linear instability of the corresponding background and prevents the usual interpretation of these estimates as predictions of a viable inflationary model. Theorem~1 establishes the existence of at least one ghost for every $\gamma\neq0$ at regular rolling background points satisfying its stated non-degeneracy conditions. Under the additional assumptions of Theorem~2, exactly one of the three scalar kinetic eigenvalues is negative. These conclusions hold for every finite $k\neq0$ within the corresponding domains of validity. The result obtained here for the cosmological branch considered throughout this work is consistent with the argument that theories of the form $f(Q,\chi,(\partial\chi)^2)$, which include $f(Q,\mathcal{L}_m)$ with a scalar inflaton, are generically ghostly \citep{jimenez2026nonviability}. In the language of that argument, coupling $\mathcal{L}_m$ to $Q$ makes the coefficient of $Q$ field-dependent and thereby revives the St\"uckelberg fields of the connection, in a manner analogous to nonlinear $f(Q)$ gravity.

\section{Discussion}\label{sec:discussion}

Our results lead to the following conclusions, which apply at different levels of generality.

\textbf{Exact result for the entire separable class:} For $f=-\alpha Q+g(\mathcal{L}_m)$, the action is off-shell equivalent, up to a boundary term, to GR minimally coupled to a $k$-essence scalar, as shown in Eq.~\eqref{eq:sepGR}. This mapping is exact for the entire separable family considered here and introduces no additional gravitational degrees of freedom beyond those of GR. For $\alpha>0$, the gravitational sector has the standard healthy sign, and the stability conditions of the scalar sector are precisely those of the corresponding $k$-essence theory. Consequently, the observational predictions are those of $k$-inflation, with the scalar Lagrangian modified by the function $g(\mathcal{L}_m)$, rather than predictions of a genuinely modified gravitational sector.

\textbf{Rigorous result for representative non-separable couplings:} For the representative interaction $\gamma Q\mathcal{L}_m$ on the trivial cosmological branch, our analysis establishes the existence of at least one scalar ghost at regular rolling background points satisfying the hypotheses of Theorem~1. These include $\rho>0$, $\mathcal{B}\neq0$, $3\gamma H^2\neq1$, $\mathcal{D}\neq0$, $k\neq0$, $\dot{\phi}\neq0$, and $(\mathcal{A}\mathcal{T},H\mathcal{U})\neq(0,0)$. Under the additional assumptions of Theorem~2, the kinetic matrix is non-degenerate and exactly one of its three eigenvalues is negative. The closed-form determinants obtained for the $-\gamma QV$ and $\gamma QX$ interactions likewise reveal kinetic instabilities under the regularity and sign conditions specified in the corresponding analysis. We emphasize that these results establish instabilities for the representative interactions and backgrounds studied here; they do not constitute an analytical proof for every possible non-separable function $f(Q,\mathcal{L}_m)$.

\begin{figure*}
\centering
\includegraphics[width=\linewidth]{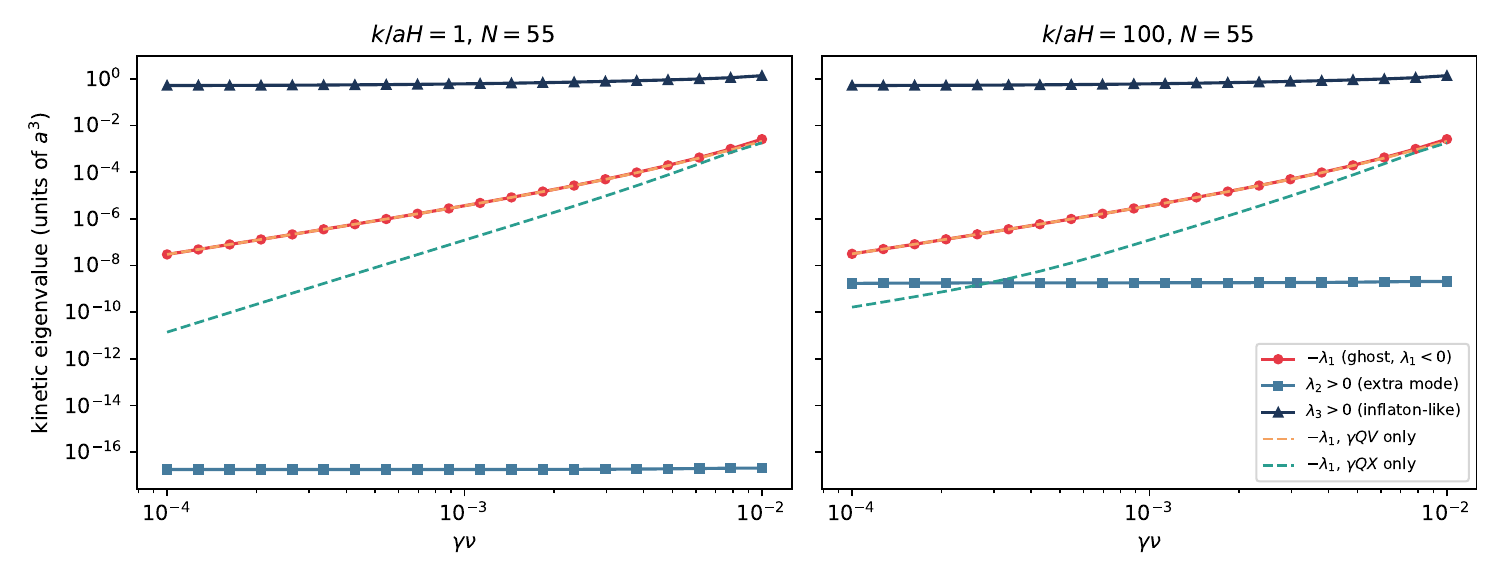}
\caption{Eigenvalues of the reduced scalar kinetic
matrix $\mathcal{K}$ in the field basis
$(\psi,E,\delta\phi)$ for the coupled model
\eqref{eq:coupled} on slow-roll backgrounds with
$V=\nu\phi^2$, $\alpha=1$, $\nu=10^{-10}$ and
$N=55$, at $k=aH$ (left) and $k=100\,aH$ (right).
Red: magnitude of the negative eigenvalue (ghost).
Blue squares: positive eigenvalue associated with the
second additional mode; its $k^4$ scaling reflects
the normalization of $E$ and disappears when using
$\tilde E=k^2E$. Dark triangles: the inflaton-like
mode. Dashed curves: negative eigenvalue for the
$-\gamma QV$ (orange, coinciding with red) and
$\gamma QX$ (teal) pieces of the coupling. The
eigenvalue magnitudes depend on the field
normalization, whereas their signs do not. At the
regular points shown, there is exactly one negative
eigenvalue, in agreement with Theorem~2.}
\label{fig:ghost}
\end{figure*}

\textbf{Conclusions supported by general arguments, but not proven here:} Theories of the form $f(Q,\chi,(\partial\chi)^2)$, which include $f(Q,\mathcal{L}_m)$ for a scalar inflaton, have been argued to be generically ghostly \citep{jimenez2026nonviability}. Moreover, it has been argued that cosmological branches of nonlinear $f(Q)$ gravity are pathological, through strong coupling or ghost instabilities \citep{gomes2024pathological}. Our explicit results are consistent with these arguments. However, we have not independently established their full generality, nor have we analyzed other functional forms or non-trivial cosmological branches.

For a canonical scalar field with $\mathcal{L}_m=X-V$, the trace of the energy--momentum tensor is $T=2X-4V$. Therefore, separable $f(Q,T)$ models that are linear in $Q$ can likewise be recast as GR with a modified scalar Lagrangian \citep{xu2019fq}. Non-separable couplings between $Q$ and $T$ fall within the broader class of field-dependent non-metricity couplings discussed in Ref.~\citep{jimenez2026nonviability}; their stability, however, must be assessed for each specific model rather than inferred solely from the present calculation.

A practical lesson concerns background-only analyses. For the coupled model, two plausible-looking estimates of $n_s$ and $r$, constructed from the same background evolution, can lead to different conclusions, as illustrated in Fig.~\ref{fig:bg}. In theories with a time-dependent $f_Q$, the modified Friedmann equations alone are insufficient to determine the physical scalar and tensor perturbation spectra. The quadratic action must be derived and its kinetic and gradient sectors examined before perturbation observables can be compared with data. In particular, the degree-of-freedom and stability analysis must precede any observational interpretation.

Our analysis has several limitations. (i) We have considered only the trivial cosmological branch, $Q=6H^2$. (ii) We have studied linear perturbations only. Whether the additional modes become strongly coupled, particularly in the limit $\gamma\to0$, requires an analysis of the canonically normalized nonlinear interactions, including at least the cubic action; this is beyond the scope of the present work. (iii) Theorems~1 and 2 exclude degenerate surfaces of the quadratic action. The relevant background-dependent quantities were checked numerically along the regular portions of the inflationary trajectories, rather than shown analytically to remain nonzero for all possible solutions. In particular, for $\gamma\nu\gtrsim10^{-2}$, the trajectories cross an isolated degenerate point at which the hypotheses of the theorems fail and the present linear analysis is inconclusive. (iv) The background-only estimates for the coupled model in Fig.~\ref{fig:bg} rely on the slow-roll approximation, whose consistency along the trajectories shown requires $\gamma\dot{\phi}^2<\alpha/2$; for the parameter choices considered, this restricts the relevant range approximately to $\gamma\nu\lesssim10^{-2}$. (v) The observational analysis of the separable class employs leading-order slow-roll expressions, assumes $N\in[50,60]$ without specifying an explicit reheating history, and uses published confidence contours rather than a full likelihood analysis. The resulting parameter intervals should therefore be regarded as indicative rather than precision constraints. Nevertheless, if the lower bound $r>0.066$ obtained in our analysis is maintained over the full parameter range considered, it lies above the commonly quoted BK18 upper limit and strongly disfavors the corresponding separable model. A definitive statistical exclusion would, however, require a comparison using the same pivot scale, data combination, and confidence level, preferably through a consistent likelihood analysis.


\section{Conclusions}\label{sec:conclusions}

In this work, we have revisited slow-roll inflation in $f(Q,\mathcal{L}_m)$ gravity with a scalar inflaton, for which the matter Lagrangian is $\mathcal{L}_m=X-V$, in order to assess whether matter--geometry couplings of this type can provide a viable modification of inflationary dynamics and observational predictions.

For the entire separable class linear in $Q$, $f=-\alpha Q+g(\mathcal{L}_m)$, we have shown that the action is off-shell equivalent, up to a boundary term, to general relativity coupled to a $k$-essence field, as expressed in Eq.~\eqref{eq:sepGR}. Consequently, this class introduces no additional gravitational degrees of freedom beyond those of standard GR. For $\alpha>0$, the gravitational sector has the standard healthy sign, while the stability of the scalar sector is entirely governed by the corresponding $k$-essence conditions given in Eq.~\eqref{eq:stability}. Thus, the inflationary predictions of these models are precisely those of the corresponding $k$-inflation models in GR. This exact mapping also clarifies the apparent sign inversion between the linear and nonlinear models discussed in the earlier literature. In the nonlinear case, one sign of the gravitational coupling leads to a ghostlike inflaton, whereas the physically admissible branch requires the opposite sign. The resulting sound speed can be slightly superluminal in the parameter range considered; this feature should be distinguished from the ghost instability and assessed separately in interpreting the effective scalar theory.

For the explicit linear and nonlinear models considered in the literature, a quadratic inflaton potential leads to an effective quadratic potential supplemented by a constant offset, with the inflationary predictions controlled by the effective parameter combination $x\equiv\beta/(\alpha\nu)$. Although these models can marginally accommodate the Planck 2018 constraints alone within a restricted parameter range, their predictions are in tension with the joint Planck, BAO, and BICEP/Keck 2018 constraints. In particular, within the range $N\leq60$ considered here, we obtain the lower bound $r>0.066$, which exceeds the commonly quoted BK18 upper bound $r<0.036$, provided that both quantities are compared at the same pivot scale and with consistent definitions. Moreover, in the parameter region where a reduction of $r$ is obtained, the required negative constant contribution to the effective potential can lead to an early termination of the post-inflationary expansion in the background solutions studied here. This raises an additional difficulty for obtaining a physically acceptable post-inflationary evolution and should be taken into account when assessing the viability of these models.

For the representative non-separable model $f=-\alpha Q+2\mathcal{L}_m+\gamma Q\mathcal{L}_m$, whose interaction contains both $\gamma QX$ and $-\gamma QV$ contributions, the effective tensor kinetic coefficient becomes time-dependent and modifies the background evolution. More importantly, our quadratic perturbation analysis in the coincident gauge, on the trivial cosmological branch, shows that the scalar kinetic matrix is generically of full rank at regular points where its determinant is nonzero. Thus, in this regime, two additional scalar modes are present beyond the usual inflaton mode. Theorem~1 establishes the existence of at least one scalar ghost whenever its stated non-degeneracy conditions are satisfied. Under the additional assumptions of Theorem~2, exactly one of the three scalar kinetic eigenvalues is negative. In the decoupling limit $\gamma\to0$, the kinetic contributions associated with the additional modes vanish and the rank of the kinetic matrix drops discontinuously. This singular change in the quadratic kinetic structure indicates that the limit requires further investigation; however, establishing whether it corresponds to strong coupling requires an analysis of the canonically normalized nonlinear interactions, which is left for future work.

Taken together, our results establish a clear distinction between the separable and representative non-separable models examined here. The entire separable class linear in $Q$ is dynamically equivalent to $k$-inflation in GR and therefore does not provide genuinely new gravitational degrees of freedom. By contrast, the representative non-separable interactions studied in this work exhibit scalar ghost instabilities at regular rolling background points within the domains specified by our theorems. These findings demonstrate why modified Friedmann equations alone cannot establish the viability of a theory when $f_Q$ depends dynamically on the matter sector. A reliable assessment requires a prior analysis of the propagating degrees of freedom and the kinetic and gradient stability of the perturbations before any comparison with cosmological observations is attempted.

The scope of these conclusions is subject to several limitations. Our perturbative instability results apply directly to the representative coupling functions and the trivial cosmological branch considered here, and to regular rolling backgrounds satisfying the hypotheses of the corresponding theorems. They do not constitute an exhaustive analytical no-go theorem for every possible functional form of $f(Q,\mathcal{L}_m)$ or for all cosmological branches. In addition, the numerical analysis identifies isolated degenerate points along some of the considered trajectories, where the assumptions underlying the perturbative theorems cease to hold and the present linear analysis becomes inconclusive. Further investigations of non-trivial connection branches, the canonically normalized cubic and higher-order actions, and broader classes of matter--geometry interactions are therefore necessary to determine whether viable non-separable extensions can be constructed.

\begin{acknowledgments}
M.S. and J.T.F. thank the Research Council of K. N. Toosi University of Technology. This work is partially supported by the Iran National Science Foundation (INSF). S.F. is supported by the Generalitat Valenciana and the European Social Fund through the APOSTD 2025 fellowship (CIAPOS/2024/461), and partially by MICIU/AEI and FEDER/EU (grant PID2024-159689NB-C21) and the EU Horizon Europe programme (NewFunFiCO, grant No. 101086251).
\end{acknowledgments}

\bibliographystyle{apsrev4-1}
\bibliography{main}


\appendix

\section{Analytic solutions for the quadratic potential}
\label{app:lambert}

For $V=\nu\phi^2$, the slow-roll expressions of Sec.~\ref{sec:separable} can be inverted analytically in terms of the Lambert $W$ function, $W_k(z)$, which satisfies $W_k(z)e^{W_k(z)}=z$. The branch choice $k$ is determined by the physical requirements that the inflaton field be real and that $H^2>0$.

\paragraph{Linear model.}
The end of inflation, defined by $\epsilon=1$, occurs at
\begin{equation}
\phi_{\mathrm{end}}^2=
\frac{2\alpha\nu+\beta+2\sqrt{\alpha\nu(\alpha\nu+\beta)}}{2\nu}.
\end{equation}
The number of $e$-folds is given by
\begin{equation}
N=\frac{1}{4\alpha\nu}
\left[\nu\phi^2-\beta\ln\phi\right]_{\phi_{\mathrm{end}}}^{\phi}.
\end{equation}
Inverting this relation yields
\begin{equation}
\phi^2=-\frac{\beta}{2\nu}\mathcal{W}_{\mathrm{L}},
\qquad
\mathcal{W}_{\mathrm{L}}\equiv W_k(z_{\mathrm{L}}),
\label{eq:WL}
\end{equation}
where
\begin{equation}
z_{\mathrm{L}}\equiv-\frac{2\nu}{\beta}e^{-A},
\end{equation}
and
\begin{equation}
\begin{aligned}
A=\frac{1}{\beta}\Bigg\{&
2\alpha\nu(4N+1)+\beta
+2\sqrt{\alpha\nu(\alpha\nu+\beta)}
\\
&-\beta\ln\left[
\frac{2\alpha\nu+\beta+2\sqrt{\alpha\nu(\alpha\nu+\beta)}}{2\nu}
\right]
\Bigg\}.
\end{aligned}
\end{equation}

For $\beta>0$, the argument $z_{\mathrm{L}}$ is negative, and the lower real branch $k=-1$, for which $\mathcal{W}_{\mathrm{L}}<-1$, must be selected to ensure $\nu\phi^2>\beta/2$ and hence $H^2>0$. The principal branch $k=0$ does not satisfy this physical condition. The reality of the lower branch requires
\begin{equation}
-\frac{1}{e}\leq z_{\mathrm{L}}<0.
\end{equation}
At the endpoint $z_{\mathrm{L}}=-1/e$, one has $\mathcal{W}_{\mathrm{L}}=-1$, which gives $H^2=0$. This endpoint is therefore excluded from the physical domain, leaving
\begin{equation}
-\frac{1}{e}<z_{\mathrm{L}}<0.
\end{equation}

Conversely, for $\beta<0$, the argument $z_{\mathrm{L}}$ is positive, and the principal branch $k=0$ is the unique real solution. In this case, inflation ends only if $|\beta|<\alpha\nu$; otherwise, $\epsilon<1$ throughout. The inflationary observables then take the closed forms
\begin{align}
n_s={}&1+
\frac{8\alpha\nu(2\mathcal{W}_{\mathrm{L}}-1)}
{\beta(\mathcal{W}_{\mathrm{L}}+1)^2},
\\
r={}&-\frac{64\alpha\nu\mathcal{W}_{\mathrm{L}}}
{\beta(\mathcal{W}_{\mathrm{L}}+1)^2}.
\end{align}

\paragraph{Nonlinear model.}
For the nonlinear model with $\alpha<0$ and $\beta<0$, the condition $\epsilon=1$ gives
\begin{equation}
\phi_{\mathrm{end}}^4=
\frac{-4\alpha\nu-\beta
+2\sqrt{2}\sqrt{\alpha\nu(2\alpha\nu+\beta)}}{4\nu^2}.
\end{equation}
The number of $e$-folds is
\begin{equation}
N=-\frac{1}{4\alpha\nu}
\left[\nu^2\phi^4+\beta\ln\phi\right]_{\phi_{\mathrm{end}}}^{\phi}.
\end{equation}
Inverting this relation gives
\begin{equation}
\phi^4=\frac{\beta}{4\nu^2}\mathcal{W}_{\mathrm{NL}},
\qquad
\mathcal{W}_{\mathrm{NL}}\equiv W_{-1}(z_{\mathrm{NL}}),
\label{eq:WNL}
\end{equation}
where
\begin{equation}
z_{\mathrm{NL}}\equiv\frac{4\nu^2}{\beta}e^B,
\end{equation}
and
\begin{equation}
\begin{aligned}
B=\frac{1}{\beta}\Bigg[
&-16\alpha\nu N-4\alpha\nu-\beta
+2\sqrt{2}\sqrt{\alpha\nu(2\alpha\nu+\beta)}
\\
&+\beta\ln\phi_{\mathrm{end}}^4
\Bigg].
\end{aligned}
\end{equation}

Since $\beta<0$, the argument $z_{\mathrm{NL}}$ is negative. The lower real branch $W_{-1}$ is selected to ensure $\phi^4>0$ and
\begin{equation}
4\nu^2\phi^4+\beta
=\beta(1+\mathcal{W}_{\mathrm{NL}})>0,
\end{equation}
which is the required positivity condition for $H^2$ in this model. The reality of this branch requires
\begin{equation}
-\frac{1}{e}\leq z_{\mathrm{NL}}<0.
\end{equation}
At the endpoint $z_{\mathrm{NL}}=-1/e$, one has $\mathcal{W}_{\mathrm{NL}}=-1$, giving $H^2=0$. Thus, the physical domain is restricted to
\begin{equation}
-\frac{1}{e}<z_{\mathrm{NL}}<0.
\end{equation}
The corresponding inflationary observables are
\begin{align}
n_s={}&1+
\frac{16\alpha\nu(2\mathcal{W}_{\mathrm{NL}}-1)}
{\beta(\mathcal{W}_{\mathrm{NL}}+1)^2},
\\
r={}&-\frac{128\alpha\nu\mathcal{W}_{\mathrm{NL}}}
{\beta(\mathcal{W}_{\mathrm{NL}}+1)^2}.
\end{align}

These analytic expressions agree with direct numerical integrations of the exact background equations (without slow-roll approximations) to within approximately $0.05\%$ in $n_s$ and $1$--$2\%$ in $r$ for the parameter choices tested. These differences quantify the accuracy of the leading-order slow-roll expressions in those cases and should not be interpreted as uniform error bounds over the full parameter space.

\section{Details of the perturbative calculation}
\label{app:pert}

The quadratic action was computed via symbolic manipulation in \textsc{SymPy}. The documented source code used for these derivations is available upon request to ensure reproducibility. Starting from the perturbed metric in Eq.~\eqref{eq:metricpert}, where all perturbations depend on $t$ and one spatial Cartesian coordinate (sufficient for scalar modes by isotropy), we calculated $\sqrt{-g}$, the inverse metric up to second order, $Q_{\alpha\mu\nu}=\partial_\alpha g_{\mu\nu}$, and the non-metricity scalar $Q$ from Eq.~\eqref{eq:Qdef}. We then expanded $\frac12\sqrt{-g}f(Q,\Lm)$ to second order.

In the coincident gauge, $\xi^\lambda=x^\lambda$, the coordinate freedom associated with the Stückelberg fields is fixed. The scalar metric perturbations and the inflaton fluctuation remain as variables in the gauge-fixed action, but not all of them represent propagating degrees of freedom. In particular, the lapse and shift perturbations, $\Phi$ and $B$, are non-dynamical and must be eliminated before identifying the propagating scalar modes. Their equations of motion contain no time derivatives, and $B$ appears linearly in the auxiliary quadratic block, for which $M_{BB}=0$. We eliminated $(\Phi,B)$ algebraically via the Schur complement of their Hessian block, wherever that block is invertible. The reduced kinetic matrix $\mathcal{K}$, acting on the field vector $\vec{Y}=(\psi,E,\delta\phi)^T$, takes the explicit structure
\begin{equation}
\mathcal{K}=
\begin{pmatrix}
\mathcal{K}_{\psi\psi} & \mathcal{K}_{\psi E} & \mathcal{K}_{\psi\delta\phi}
\\
\mathcal{K}_{E\psi} & 0 & \mathcal{K}_{E\delta\phi}
\\
\mathcal{K}_{\delta\phi\psi} & \mathcal{K}_{\delta\phi E} & \mathcal{K}_{\delta\phi\delta\phi}
\end{pmatrix},
\end{equation}
where $\mathcal{K}_{EE}=0$ holds identically for general $\alpha$, $a$, and $k$, and all non-vanishing entries are rational functions of background quantities scaled by the governing polynomials. We explicitly verified that
\begin{enumerate}
\item[(i)] The zeroth-order Lagrangian reduces identically to the minisuperspace Lagrangian yielding Eqs.~\eqref{eq:cF1}--\eqref{eq:ckg}.

\item[(ii)] All first-order (tadpole) terms vanish upon imposing the background equations of motion.

\item[(iii)] For $\gamma=0$, as well as for the separable model $f=-\alpha Q+(2\Lm)^2$, the reduced kinetic matrix correctly drops to rank one.

\item[(iv)] For $f=-\alpha Q+\lambda Q^2+2\Lm$, two additional scalar modes propagate, in full agreement with Ref.~\citep{jimenez2020cosmology}.

\item[(v)] The tensor mass term vanishes on shell, recovering Eq.~\eqref{eq:tensor}.
\end{enumerate}

Furthermore, the analytical determinant \eqref{eq:detQL} was independently cross-checked against a numerical Schur complement computed at 30 random background points (with general $\alpha$, $a$, $k$, and both signs of $\gamma$) using 60-digit arbitrary-precision arithmetic, achieving a relative agreement of $10^{-46}$.

The auxiliary sub-block satisfies $M_{BB}=0$ and
\begin{equation}
M_{\Phi B}=
\frac{a^2k^2\mathcal{D}}
{4(3\gamma H^2-1)\mathcal{B}},
\end{equation}
leading to
\begin{equation}
\det M_{aa}=-M_{\Phi B}^2.
\end{equation}
The determinant \eqref{eq:detQL} was derived using the Schur determinant formula
\begin{equation}
\det\mathcal{K}
=
\frac{
\det\begin{pmatrix}
M_{vv} & M_{va}\\
M_{av} & M_{aa}
\end{pmatrix}
}{
\det M_{aa}
}.
\end{equation}
In evaluating the polynomial determinant, the denominators were cleared row by row by multiplying each row by the corresponding least common multiple of its denominators. The associated overall factors were then restored when reconstructing the original determinant. This procedure allows the polynomial part of the calculation to be performed exactly without losing denominator-dependent contributions. All entries of $\mathcal{K}$ scale as $a^3$, with those in the $E$ row and column scaling as $k^2$.

The polynomials appearing in Eqs.~\eqref{eq:Koff} and \eqref{eq:detQL} are given by
\begin{align}
\mathcal{D}={}&
243\gamma^5H^7\dot\phi^4
+54\gamma^5H^6V'\dot\phi^3
-162\alpha\gamma^3H^5\dot\phi^2
\nonumber\\
&+189\gamma^4H^5\dot\phi^4
-18\alpha\gamma^3H^4V'\dot\phi
+18\gamma^4H^4V'\dot\phi^3
\nonumber\\
&+12\alpha^2\gamma H^3
-24\alpha\gamma^2H^3\dot\phi^2
+9\gamma^3H^3\dot\phi^4
\nonumber\\
&-6\gamma^3H^2V'\dot\phi^3
+4\alpha^2H
-6\alpha\gamma H\dot\phi^2
-\gamma^2H\dot\phi^4
\nonumber\\
&+2\alpha\gamma V'\dot\phi
-2\gamma^2V'\dot\phi^3,
\\[2pt]
\mathcal{T}={}&
81\gamma^3H^5\dot\phi^3
+18\gamma^3H^4V'\dot\phi^2
-18\alpha\gamma H^3\dot\phi
\nonumber\\
&+18\gamma^2H^3\dot\phi^3
-6\alpha\gamma H^2V'
+12\gamma^2H^2V'\dot\phi^2
\nonumber\\
&-6\alpha H\dot\phi
+9\gamma H\dot\phi^3
-2\alpha V'
+2\gamma V'\dot\phi^2,
\\[2pt]
\mathcal{U}={}&
243\gamma^5H^7\dot\phi^4
+54\gamma^5H^6V'\dot\phi^3
-135\alpha\gamma^3H^5\dot\phi^2
\nonumber\\
&+135\gamma^4H^5\dot\phi^4
-18\alpha\gamma^3H^4V'\dot\phi
+18\gamma^4H^4V'\dot\phi^3
\nonumber\\
&+9\alpha^2\gamma H^3
-18\alpha\gamma^2H^3\dot\phi^2
+9\gamma^3H^3\dot\phi^4
\nonumber\\
&-6\gamma^3H^2V'\dot\phi^3
+3\alpha^2H
-3\alpha\gamma H\dot\phi^2
-3\gamma^2H\dot\phi^4
\nonumber\\
&+2\alpha\gamma V'\dot\phi
-2\gamma^2V'\dot\phi^3.
\end{align}

The polynomials in Eq.~\eqref{eq:detQX} are
\begin{align}
\mathcal{S}_1={}&
27\gamma^2H^3\dot\phi^3
-18\gamma H^3\dot\phi
+9\gamma H\dot\phi^3
\nonumber\\
&-6H\dot\phi
+\gamma V'\dot\phi^2
-2V',
\\[2pt]
\mathcal{S}_2={}&
54\gamma^3H^3\dot\phi^4
-36\gamma^2H^3\dot\phi^2
-24\gamma H^3
\nonumber\\
&+6\gamma^2H\dot\phi^4
+4\gamma H\dot\phi^2
-8H
\nonumber\\
&+\gamma^2V'\dot\phi^3
-2\gamma V'\dot\phi.
\end{align}

For the numerical eigenvalues shown in Fig.~\ref{fig:ghost}, background quantities were extracted from the slow-roll solutions of Eqs.~\eqref{eq:cF1}--\eqref{eq:ckg}, with $\dot H$ and $\ddot\phi$ evaluated via the exact equations of motion, and eigenvalues were determined with 60-digit arbitrary-precision arithmetic.

\end{document}